\documentclass[a4paper,fleqn,usenatbib]{mnras}

\usepackage{newtxtext,newtxmath}

\usepackage[T1]{fontenc}
\usepackage{ae,aecompl}

\usepackage{graphicx}	
\usepackage{amsmath}	
\usepackage{amssymb}	

\title[]{Absolute colors and phase coefficients of trans-Neptunian objects: $H_{V}-H_{R}$ and relative phase coefficients}

\author[C. Ayala-Loera et al.]
{C. Ayala-Loera,$^{1}$\thanks{E-mail: madelcarmen@on.br}
A.~Alvarez-Candal,$^{1}$
J. L.~Ortiz,$^{2}$
R.~Duffard,$^{2}$
\newauthor{E.~Fern\'andez-Valenzuela,$^{2,3}$
P.~Santos-Sanz,$^{2}$
and N.~Morales,$^{2}$
}
\\
$^{1}$Observat\'orio Nacional / MCTIC, Rua General Jos\'e Cristino 77, Rio de Janeiro, RJ, 20921-400, Brazil.\\
$^{2}$Instituto de Astrof\'isica de Andaluc\'ia, CSIC, Apt 3004, 18080, Granada, Spain.\\
$^{3}$Florida Space Institute (FSI) at University of Central Florida, 02354 Research Parkway, Suite 284, Orlando, FL, 32826, USA.
}

\date{Accepted XXX. Received YYY; in original form ZZZ}

\pubyear{2018}

\begin{document}
\label{firstpage}
\pagerange{\pageref{firstpage}--\pageref{lastpage}}
\maketitle

\begin{abstract}
The trans-Neptunian objects (TNOs) are small Solar System bodies at large distances from the Sun. As such, their physical properties are difficult to measure. 
Accurate determination of their physical parameters is essential to model and theorize the actual composition and distribution of the population, 
and to improve our understanding of the formation and evolution of the Solar System. The objective of this work is to construct phase curves in two filters, 
$V$ and $R$, of a large TNO sample obtaining absolute magnitudes ($H$) and phase coefficients ($\beta$), and study possible relations between them and 
other physical parameters (orbital elements, sizes, and albedos). We used our own data, together with data from the literature, to create the phase curves 
assuming an overall linear trend. 
We obtained new magnitudes for 35 TNOs, 27 in the V filter and 35 in the R filter. These magnitudes, together with data from the literature, 
allowed us to obtain absolutes magnitudes, 114 in the V filter and 113 in the R filter, of which 106 have both. From the search for 
correlations we found a strong anticorrelation between $H_V-H_R$ and $\Delta \beta=\beta_V-\beta_R$, which is probably more related to 
surface structure than to composition or size of the objects.
\end{abstract}




\section{Introduction}

The trans-Neptunian objects, TNOs, are distant objects leftovers of the protoplanetary disk where the planets formed. 
The understanding of their physical properties sets important constrains to improve the evolution models of the Solar System \citep{mulle10}.

Nowadays, the Minor Planet Center\footnote{http://www.minorplanetcenter.net/iau/mpc.html} lists around 2,300 TNOs.  
Unfortunately, just a few hundreds of them have high-quality physical studies, due to their orbital and size distributions, 
that produce few objects brighter than $V_{mag}~\sim~17$.
Among the techniques used to study TNOs, photometry is the less expensive one (in terms of observing time). Photometric studies allow to obtain information 
of a good number of TNOs via apparent magnitudes and colors.
The first are measurements of the integral reflected light by the TNO surface, subjected to the geometry of the observation and physical properties, 
such as diameter ($D$) and albedo ($p$), while the latter is a measure of the slope of the spectral reflectance of the object.
Apparent magnitudes can be used to obtain absolute magnitudes ($H$) if the observational circumstances are known. The absolute magnitude 
is the mean apparent magnitude, over a rotation 
cycle of the object, observed at zero phase angle, and both at 1 AU from the Sun and the Earth. In practice, $H$ should be computed using phase curves 
and the formalism of \cite{muin10}. 
A phase curve shows the change of the apparent magnitude, normalized to unit distance from the Earth and the Sun, with the phase angle ($\alpha$).
Nonetheless, due to the large distances where the TNOs reside, $\alpha$ (the arc that subtends the distance Sun-Earth as seen from the object) can 
only reach values as large as $\sim3^{\rm o}$, while the centaurs (representatives of the TNO population orbiting closer to the Sun) can be observed 
up to phase angles $\sim 7^{\rm o}$. 
In these small ranges the phase curves can be approximated by a linear function \cite[see, for instance,][]{alvarez-candal16} with due caution for possible opposition 
surges at phase angles close to zero. 

The absolute magnitude is of interest because it can be used as a proxy for size through
\begin{equation}\label{eq1}
	D~[km] = C \times10^{-H/5}p^{-1/2},
\end{equation}
\noindent where $C$ is a constant.

On the other hand, colors are the difference of two magnitudes measured using two filters with different effective wavelengths, $\lambda_1$ and $\lambda_2$, and, as
mentioned, are related to the reflectance spectrum of the objects, or, in other words, to their surface composition. However, even if different colors might imply 
different compositions, they cannot be used to infer it, but just as a first approach \citep{dores08,baruc11}.

The TNOs show a great diversity of colors, ranging from neutral to very red \citep{baruc05}. It has been suggested that colors of TNOs together with other properties, 
such as sizes and albedos at different wavelengths, are used to help describe their surfaces properties and evolution \citep{Luu1996, Pike2017}. 
However, laboratory work of \citet{Kanuchova12} showed that fully weathered organic materials variate refractive properties of the materials and turn back in the 
color$-$color diagrams, such as a result suggest that colors themselves might not be entirely useful to explain the TNOs evolution. 
Thus, the color diversity of TNOs must be explained taking in account nurture and nature scenarios. 

For instance, \cite{peixe12} reported a bimodal $(B-R)$ distribution of centaurs and small TNOs with a gap in $(B-R)\sim 1.6$.
Such a bimodality was independent of their orbital distribution and could be explained by different location of origin and/or disruptive collisions processes. 
Also, \cite{lacer14} reported two groups of mid-sized TNOs based in albedo and color: one bright and red, while the other is dark and neutral, 
with no dynamical segregation. Such a color-albedo separation was explained as different birth locations and is considered as evidence of a break in the composition 
continuity of the protoplanetary disk. 
   
In a previous work \citep[hereafter paper 1]{alvarez-candal16} we analyzed the absolute magnitudes ($H_V$) and phase coefficients ($\beta_V$), of 110 TNOs in the V band. 
The methodology we used in paper 1 was slightly different than the presented here. We only used $V$ magnitudes. In cases when only $R$ magnitudes were available, 
we transformed them to $V$ magnitudes using the weighed average $(V-R)$ for the object.
In the present work we extend our analysis by including data in the $R$ band, i.e., $H_R$ and $\beta_R,$ and an updated list of magnitudes, some observed 
by ourselves and other from the literature not included before. With this, we aim at gaining a deeper comprehension on the surface characterization of TNOs. Of special interest is the 
``absolute color'', $H_V-H_R$, that is proportional to the ratio of albedos (Eq. \ref{eq1}) and does not have any phase-related effect, providing a zero-phase 
approximation to the reflectance spectrum. We also define the ``relative phase coefficient'' as $\Delta\beta=\beta_V-\beta_R$ and study its relationship with 
the absolute color and their relation with other typical parameters.

This paper is organized as follows: in the next section we described our new observations, in Sect. 3 we explain the method used, while in Sects. 4 
and 5 we present and discuss our results. 
\section{Observations and data reduction}
New observations of 35 objects are reported in this work. The observations were carried out using the 2.2-m telescope, at the Calar Alto Observatory (Spain), 
and the Southern Astrophysical Research (SOAR) Telescope, 4.1-m telescope, located at Cerro Tololo Inter-American Observatory (Chile).
The observations were carried out using the $V$ and $R$ filters, with a total exposure time of 1800 s per filter, split, typically, into $3\times600$ s. 
No differential tracking was used since the sum of shorter exposures allows us to minimize trailing on the images, for example a typical centaur can move up to about 2 arcsecs in 10 minutes, therefore, using exposure times of 300 s the length of the trail is about the typical FWHM of the images. Also, short exposures help to reduce the effects of background sources, hot pixels, and cosmic rays hits.
Standard stars from \cite{lando92} were also observed at different airmasses ($\lesssim 1.4$) to transform the instrumental 
magnitudes to the standard system. 

All images were bias and flat-field corrected in the usual way using daily calibration files and standard IRAF routines\footnote{IRAF is distributed by the National Optical Astronomy Observatory, which is operated by the Association of Universities for Research in Astronomy (AURA) under a cooperative agreement with the National Science Foundation.}.
The images of the science objects were then aligned, using \texttt{imalign}, and median combined, using \texttt{imcombine}. 
The median combination is useful as it removes hot pixels and cosmic ray hits. We used the task \texttt{phot} to do aperture photometry 
of the science object and the standard stars, using a fixed aperture to 
obtain apparent instrumental magnitudes. 
In the few cases, crowded fields for example, where it was not possible to perform aperture photometry, we used aperture correction.

We followed the procedures outlined in paper 1 to correct the 
apparent instrumental magnitudes of atmospheric extinction, to compute the zero points of the nights (or used averages when necessary), 
and to propagate the corresponding errors. Finally, we calculated the reduced magnitudes,
\begin{equation}
    M(1,1,\alpha) = M - 5 log (r \Delta),
\end{equation}
where $\Delta$ is the topocentric distance of the object, and $r$ is its heliocentric distance. Both were obtained from JPL-Horizons 
ephemeris\footnote{http://ssd.jpl.nasa.gov/horizons.cgi}.
All information is contained in Table~\ref{table:observa}.
\begin{table*}
	\centering
	\caption{Observations}
	\label{table:observa}
	\begin{tabular}{lcccccccr} 
		\hline
		 Object                &       $ V$       &    $ R $          &   Night      &  $r$ (AU) & $\Delta$ (AU)& $\alpha$ ($^\circ$)&  Telescope &  Notes \\ 
		\hline
		19308  1996 TO$_{66}$  & 20.91$\pm$0.249  &  20.17$\pm$0.176  &  2014-07-20  &  47.1403  &   46.9431   &   1.2159    &    CAHA         & 1    \\
		44594  1999 OX$_{3}$   & 20.99$\pm$0.275  &  20.06$\pm$0.102  &  2014-07-20  &  19.9503  &   19.3693   &   2.4347    &    CAHA         & 1    \\
		47932  2000 GN$_{171}$ & 21.00$\pm$0.103  &  20.71$\pm$0.080  &  2014-06-22  &  28.4215  &   27.6271   &   1.2949    &    SOAR         & 1    \\
		82158  2001 FP$_{185}$ & 22.30$\pm$0.214  &  21.56$\pm$0.091  &  2014-07-20  &  35.7917  &   35.8120   &   1.6249    &    CAHA         & 1    \\
		82158  2001 FP$_{185}$ &  $\cdots$        &  22.15$\pm$0.531  &  2014-07-18  &  35.7902  &   35.7788   &   1.6260    &    CAHA         & 1    \\
		82155  2001 FZ$_{173}$ & 21.55$\pm$0.147  &  20.95$\pm$0.112  &  2014-05-30  &  32.5456  &   31.7748   &   1.1679    &    SOAR         & 1    \\
		82155  2001 FZ$_{173}$ & 21.74$\pm$0.170  &  21.70$\pm$0.182  &  2014-06-22  &  32.5507  &   32.0636   &   1.5808    &    SOAR         & 1    \\
		       2001 KD$_{77}$  & 22.25$\pm$0.149  &  21.35$\pm$0.118  &  2014-05-30  &  36.0591  &   35.1262   &   0.6455    &    SOAR         & 1    \\
		       2001 QC$_{298}$ & 23.36$\pm$0.687  &  23.14$\pm$0.635  &  2014-07-22  &  40.7801  &   40.1979   &   1.1823    &    CAHA         &      \\
		275809 2001 QY$_{297}$ & 22.31$\pm$0.167  &  21.59$\pm$0.128  &  2014-06-22  &  43.5248  &   42.9332   &   1.0962    &    SOAR         & 1    \\
		119951 2002 KX$_{14}$  & 20.77$\pm$0.150  &  20.12$\pm$0.121  &  2014-05-30  &  39.2447  &   38.2325   &   0.0920    &    SOAR         & 1    \\
		119951 2002 KX$_{14}$  & 20.71$\pm$0.095  &  20.18$\pm$0.068  &  2014-06-22  &  39.2422  &   38.2758   &   0.4645    &    SOAR         & 1    \\
		120178 2003 OP$_{32}$  &  $\cdots$        &  19.79$\pm$0.042  &  2014-07-20  &  41.9373  &   41.1515   &   0.8933    &    CAHA         & 1    \\
		307616 2003 QW$_{90}$  & 21.09$\pm$0.324  &  20.99$\pm$0.143  &  2014-07-22  &  43.6784  &   43.3774   &   1.2798    &    CAHA         &      \\
		120216 2004 EW$_{95}$  & 21.23$\pm$0.118  &  22.08$\pm$0.372  &  2014-06-22  &  27.0915  &   26.4941   &   1.7573    &    SOAR         & 1    \\
		90568  2004 GV$_{9}$   & 20.22$\pm$0.145  &  19.61$\pm$0.111  &  2014-05-30  &  39.3563  &   38.4608   &   0.6945    &    SOAR         & 1    \\
		       2004 NT$_{33}$  & 20.73$\pm$0.107  &  20.29$\pm$0.046  &  2014-07-20  &  38.6737  &   37.9280   &   1.0360    &    CAHA         & 1    \\
		307982 2004 PG$_{115}$ & 21.20$\pm$0.365  &  20.58$\pm$0.152  &  2014-07-19  &  37.4889  &   36.6666   &   0.9276    &    CAHA         & 1    \\
		145452 2005 RN$_{43}$  & 20.03$\pm$0.137  &  19.74$\pm$0.110  &  2014-07-20  &  40.6419  &   39.8866   &   0.9721    &    CAHA         & 1    \\
		145480 2005 TB$_{190}$ & 21.18$\pm$0.247  &  20.71$\pm$0.139  &  2014-07-20  &  46.2178  &   45.5999   &   1.0111    &    CAHA         & 1    \\
		202421 2005 UQ$_{513}$ & 21.71$\pm$0.404  &  20.43$\pm$0.198  &  2014-07-20  &  48.3909  &   48.2745   &   1.1980    &    CAHA         & 1    \\
		248835 2006 SX$_{368}$ & 21.81$\pm$0.459  &  21.24$\pm$0.238  &  2014-07-20  &  12.9767  &   13.1433   &   4.3995    &    CAHA         & 1    \\
		278361 2007 JJ$_{43}$  &  $\cdots$        &  20.35$\pm$0.265  &  2014-07-17  &  41.2853  &   40.6122   &   1.0608    &    CAHA         & 1    \\
		       2007 JK$_{43}$  & 20.90$\pm$0.146  &  20.58$\pm$0.154  &  2014-07-22  &  23.6749  &   23.0767   &   2.0091    &    CAHA         &      \\
		       2007 OC$_{10}$  & 21.14$\pm$0.166  &  20.62$\pm$0.163  &  2014-07-22  &  35.6835  &   34.8029   &   0.8284    &    CAHA         &     \\
		       2008 OG$_{19}$  &  $\cdots$        &  22.13$\pm$0.385  &  2014-07-18  &  38.5786  &   37.5909   &   0.3652    &    CAHA         & 1    \\
		       2008 OG$_{19}$  &  $\cdots$        &  21.53$\pm$0.385  &  2014-07-17  &  38.5786  &   37.5939   &   0.3838    &    CAHA         & 1    \\
		       2008 OG$_{19}$  & 21.07$\pm$0.104  &  20.58$\pm$0.076  &  2014-06-22  &  38.5781  &   37.7704   &   0.9270    &    SOAR         & 1    \\
		       2008 OG$_{19}$  & 20.50$\pm$0.140  &  20.04$\pm$0.106  &  2014-05-30  &  38.5777  &   38.0398   &   1.2894    &    SOAR         & 1    \\
		       65489     Ceto  &  $\cdots$        &  20.84$\pm$0.188  &  2014-07-20  &  35.2266  &   34.9560   &   1.5970    &    CAHA         &      \\
		       65489     Ceto  & 22.06$\pm$0.354  &  22.56$\pm$0.646  &  2014-07-22  &  35.2314  &   34.9913   &   1.6089    &    CAHA         & 1    \\
		       2060    Chiron  & 18.30$\pm$0.066  &  18.07$\pm$0.040  &  2014-07-19  &  17.8793  &   17.2129   &   2.5103    &    CAHA         & 1    \\
		       2060    Chiron  &  $\cdots$        &  17.88$\pm$0.048  &  2014-07-18  &  17.8786  &   17.2239   &   2.5426    &    CAHA         & 1    \\
		       5145    Pholus  &  $\cdots$        &  21.22$\pm$0.309  &  2014-07-22  &  25.8855  &   25.1171   &   1.4889    &    CAHA         &      \\
		       120347 Salacia  & 21.24$\pm$0.245  &  20.17$\pm$0.128  &  2014-07-19  &  44.4793  &   44.0186   &   1.1759    &    CAHA         & 1    \\
		       174567   Varda  & 20.38$\pm$0.143  &  19.80$\pm$0.110  &  2014-05-30  &  47.2155  &   46.2716   &   0.4555    &    SOAR        & 1    \\
		\hline
		(1) Average ext. coeff.&                 &                   &                 &        &                 &          &                 &     \\
	\end{tabular}	
\end{table*}

\section{Methods}
As we did in paper 1, we used our own data, complemented by data collected from the literature. 
All references are given in Table \ref{table:magnitudesH}. References already provided in paper 1 are not repeated here.

In order to obtain $H$ and $\beta$ we fitted a 1st degree polynomial to the reduced magnitudes according to
   \begin{equation}\label{eq.3}
    M(1,1,\alpha) = H + \alpha \times \beta.
   \end{equation} 
The solutions to Eq. \ref{eq.3} are $H$, as the y-intercept, and $\beta$ as the slope. Note that the same equation applies for $V$ and $R$ data.
The linear fit, although simple, minimizes the number of free parameters and describes well enough the observational data, especially considering
the restricted range of $\alpha$ we are using. While fitting, the reduced magnitude was weighted by its error, which is the same error that we have 
obtained from the apparent magnitudes. 

As mentioned above, $H$ represents a magnitude averaged over a rotation cycle. Unfortunately, many of the observations reported are snapshots at 
one unknown rotational phase. We will try to overcome this shortcoming following the procedure outlined below.

We generate 100,000 solutions of Eq. \ref{eq.3} by changing the reduced magnitude according to 
   \begin{equation}\label{eq_rand}
    M(1,1,\alpha)_i = M(1,1,\alpha) + rand_i \times \Delta m,
   \end{equation}
where $\Delta m$ is the rotational light-curve amplitude (from \citealt{thiro10, thiro12, Benecchi2013}) and $rand_i$ is a random number extracted 
from an uniform distribution within the range [-1, 1]. In cases where $\Delta m$ is unknown, we assumed $\Delta m=0.14$, which is the median value of the distribution. 
Therefore, $H$ and $\beta$ will be the average over the 100,000 solutions of Eq. \ref{eq_rand} and the errors their respective standard deviations.

\begin{center}

\begin{figure}
\includegraphics[scale=0.23]{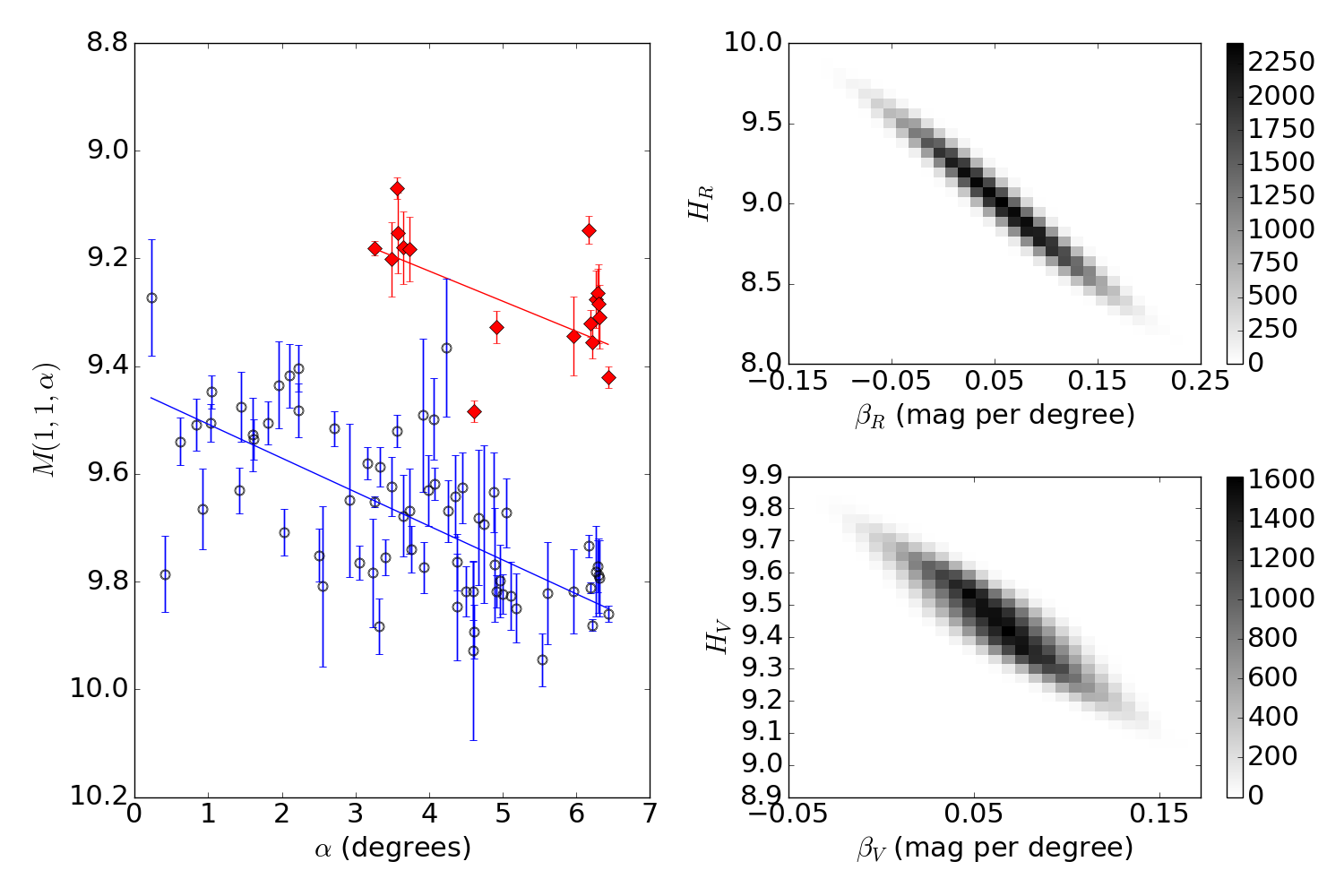}
\includegraphics[scale=0.23]{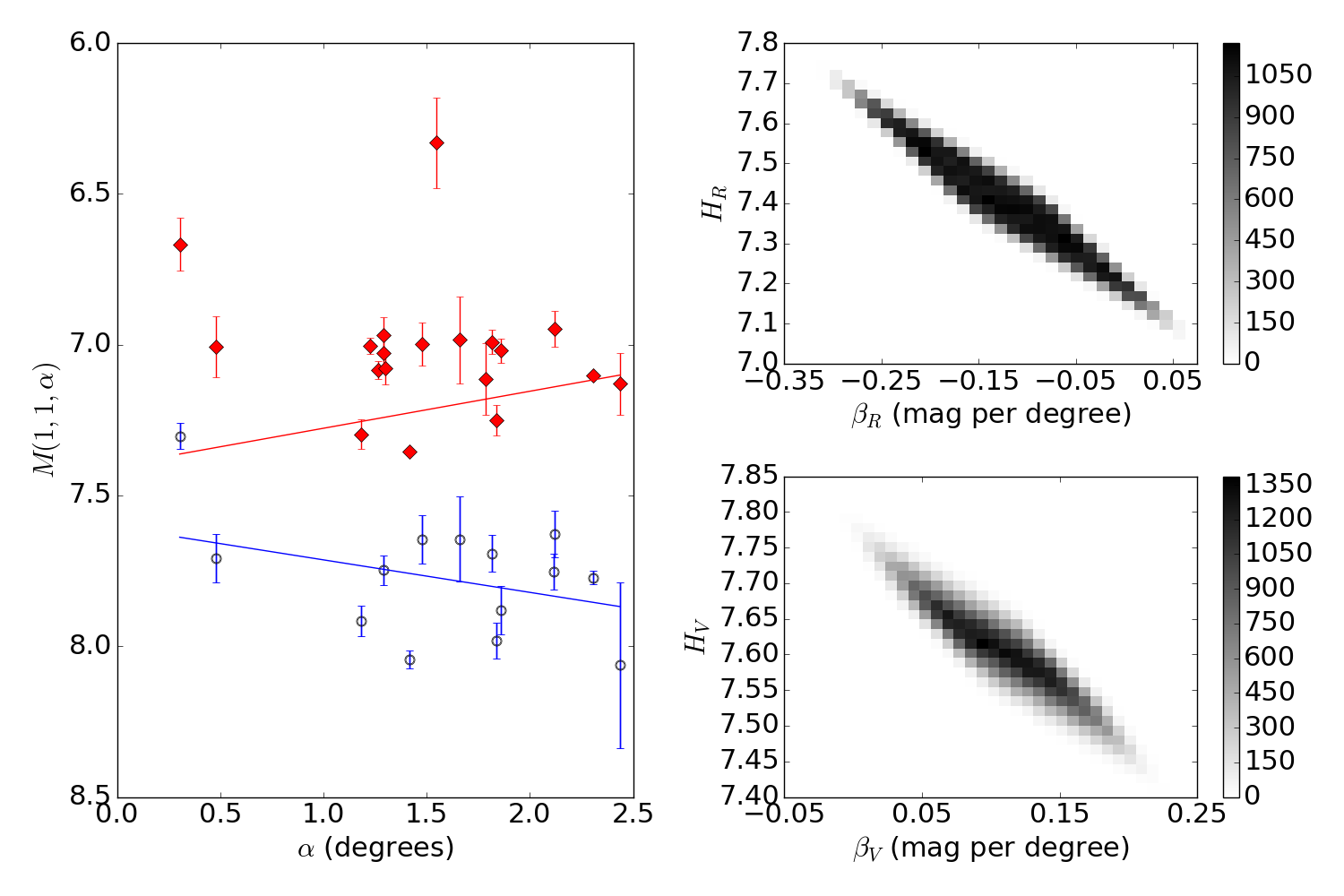}
\includegraphics[scale=0.23]{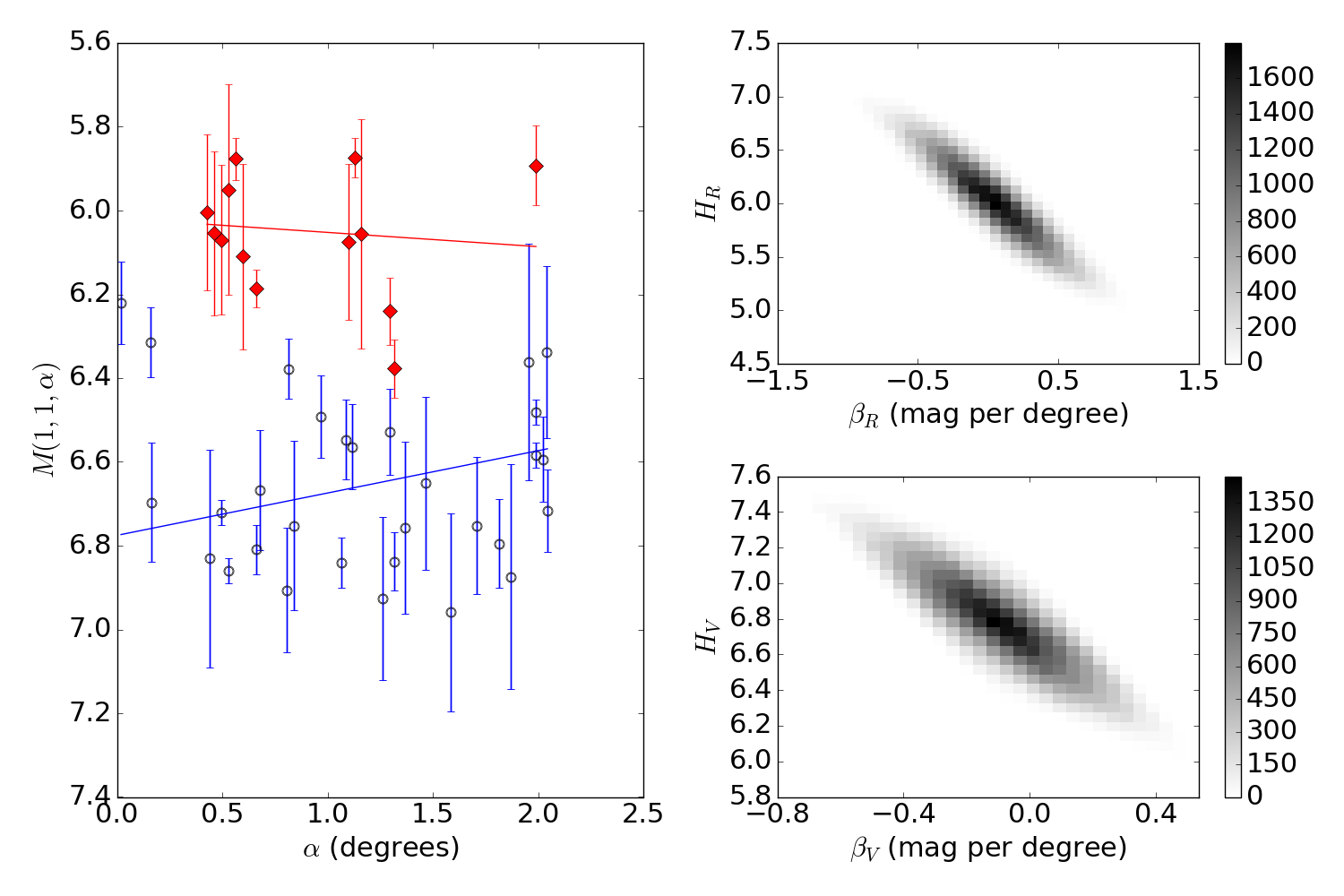}
\caption{Phase curves of Thereus (top), 1999 OX$_3$ (middle) and 2000 GN$_{171}$ (bottom). The blue points correspond to $V(1,1,\alpha)$, while 
the red diamonds indicate $R(1,1,\alpha)$.
The solid lines show our preferred solution to Eq.\ref{eq_rand}. The density plots show the phase space covered by the 100,000 solutions (see text).}
\label{fig:figura1}
\end{figure}
\end{center}

Examples are shown in Fig. \ref{fig:figura1}, where the panels in the left column show the observational data and the best fits. 
The right column shows the phase space covered by the 100,000 solutions, for data in both filters. 

Most of the objects follow the expected behavior: brighter with decreasing phase angle, implying positive values of $\beta$. 
But, there are some objects that present a peculiar behavior, that is of a decrease of magnitude with decreasing phase angle, as seen in the middle and bottom 
panels of Fig. \ref{fig:figura1}. Nevertheless, note that the phase space in these cases could allow solutions with positive values of $\beta$.

We obtained $H_V$ for 114 objects and $H_R$ for 113 objects, 105 objects have both. Figures for all our sample can be downloaded from 
{\tt http://extranet.on.br/alvarez/TNOs-Abs\_Mags/phase-curves.tar}.
\subsection{Distributions}                                                                     
\begin{center}
\begin{figure}
\includegraphics[scale=0.19]{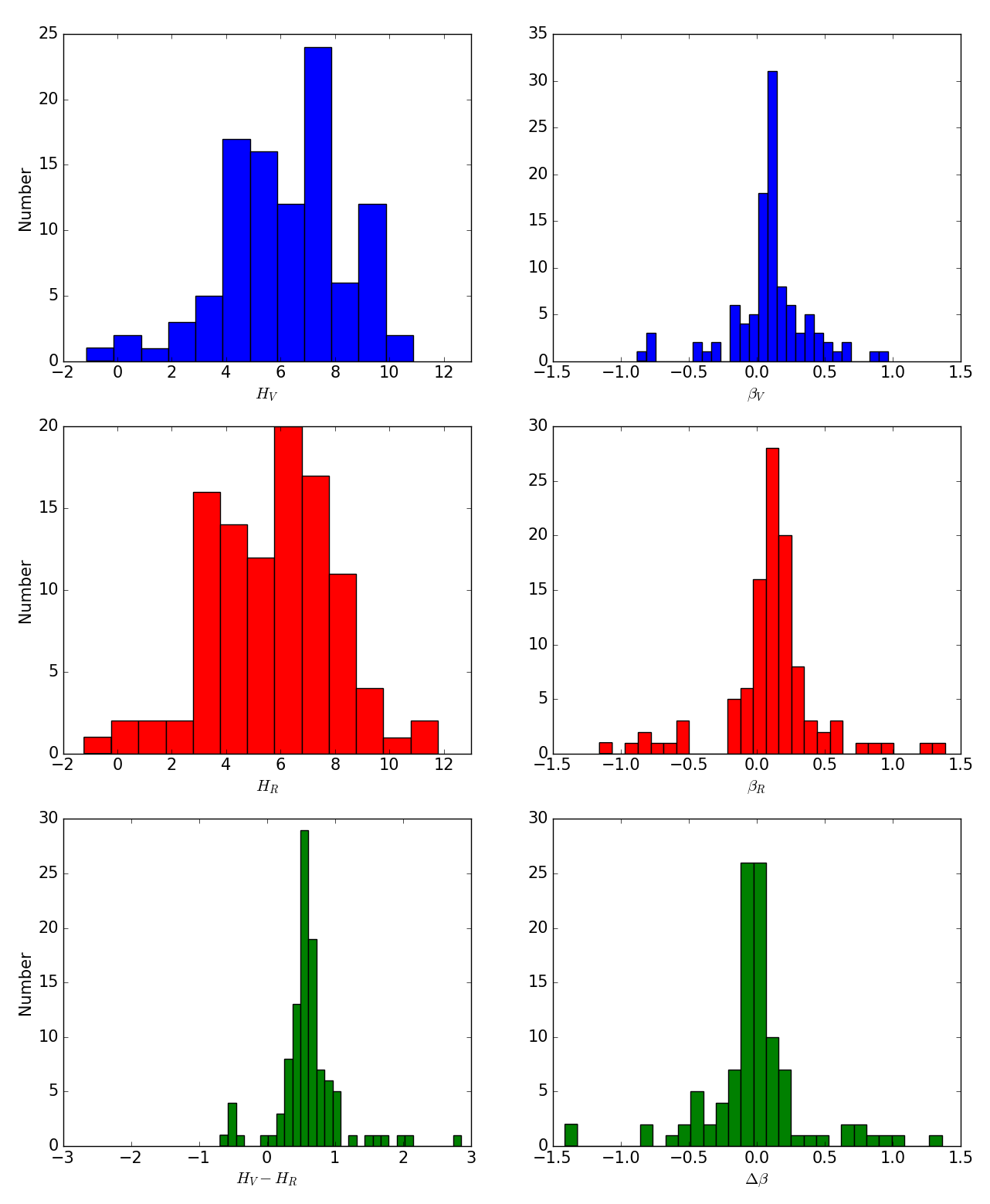}
\caption{The histograms show the distributions of $H_V$ and $\beta_V$ (top), 
and $H_R$ and $\beta_{R}$ (bottom) of Sample [1].}
\label{fig:figura2}
\end{figure}
\end{center}

The distribution of the results, absolute magnitudes, phase coefficients, $H_V-H_R$, and $\Delta\beta$ obtained for our sample are shown in Fig. \ref{fig:figura2}.
The minimum, average, and maximum of each distribution are reported in Table~\ref{table:histo}.
\begin{table}
\centering 
\caption{Notable values of the distributions}\label{table:histo}  
\begin{tabular}{|c|c|c|c|}
\hline 
  Quantity            & $Min$    &  $Max$   &  $Mean$  \\  
\hline                                                        
         $H_V$      & -1.128   & 11.806   &   6.396  \\    
   $\beta_{V}$   & -1.137   & 0.9676   &   0.079  \\  
        $H_R $       & -1.224   & 12.28   &    5.782  \\    
   $\beta_{R}$   & -1.420   &  1.384   &   0.094  \\     
$\Delta \beta$   & -1.409   & 1.3614   &  -0.014  \\     
$H_V-H_R$        & -0.689   &  2.853   &   0.598  \\     
\hline                                                              
\end{tabular}   
\end{table}  
At first glance, $H_V$ (top left panel) and $H_R$ (middle left panel) have similar overall distributions, with some differences in the detail. Likewise for $\beta_V$ (top right) and $\beta_R$ (middle right). In these last cases, most of the objects
have positive values of the phase coefficient, with a clear maximum at about 0.02 mag per degree. Nonetheless, it is clear that negative values, even as large as -1 mag per degree, are possible (see discussion below).

Most absolute colors are red, with a large concentration at $H_V-H_R\sim0.6$, while $\Delta\beta=0$ is the clear mode of its distribution, which is fairly symmetrical.


     \subsection{Search for correlations}
We test our results ($H_V$, $H_R$, $H_V-H_R$, $\beta_V$, $\beta_R$, and $\Delta\beta$) for correlations among themselves and
against other usual parameters, such as sizes, albedos, and orbital elements. We used the Spearman Correlation test that checks for 
the linear dependence between two ranked variables. It assesses monotonic relationships, whether linear or not. 

The Spearman Correlation test provides two values, the coefficient $r_s$ and $P_{r_s}$. 
The first has the form:
\begin{equation}
    r_s = 1 - \frac{6}{n(n^2-1)} \sum_{i=1}^n d_{i}^2,
\end{equation}
\noindent where $d_i$ is the difference of the assigned ranges between two variables, $x_i$ and $y_i$, and $n$ is the number of assigned pairs of data. 
Two pair of variables are correlated if $|r_s|\rightarrow1$, while if $|r_s|\approx0$ no correlation 
exists. The null hypothesis, that the two pair of variables are not correlated, is tested with $P_{r_s}$. In practice, the null hypothesis 
could be rejected if $P_{r_s}$ tends to zero. In this work, we consider a correlation as significant if $|r_s|>0.5$ and $P_{r_s}<0.0015$ (significance 
over $3\sigma$).

Our data includes all orbital sub-populations, from centaurs to detached objects \cite[see][for definitions]{gladman08}. But for 
the sake of this work we will not analyze them separated as we consider that splitting our sample into smaller ones will only decrease its statistical reliability. Nevertheless, a small discussion is included in Sect. \ref{sec4}.

All the correlation results are reported in Table \ref{table:correlations}, where we marked  in boldface those that are statistically significant. 
Among them the correlation between $H_V$ (or $H_R$) and $D$ (or $p_V$), should be simply explained by the relation between diameter, 
albedo, and absolute magnitude (Eq. \ref{eq1}).

As can be seen in Table \ref{table:correlations}, we used three separated
samples: Sample [1], includes all objects in our database;
Sample [2], includes only objects with H$_V$ fainter than 4.5;
and Sample [3], includes only objects with H$_V$ brighter than
4.5 (see below for the justification).

\noindent
\textbf{Sample [1]}:\\
\noindent
It contains 114 objects with $H_V$ and $\beta_V$, 113 with $H_R$ and $\beta_R$, and 105 in common. Among them 64 objects 
have albedos and diameters reported by the {\em TNOs are cool} survey\footnote{http://public-tnosarecool.lesia.obspm.fr}.

\begin{figure}
\centering
\includegraphics[scale=0.45]{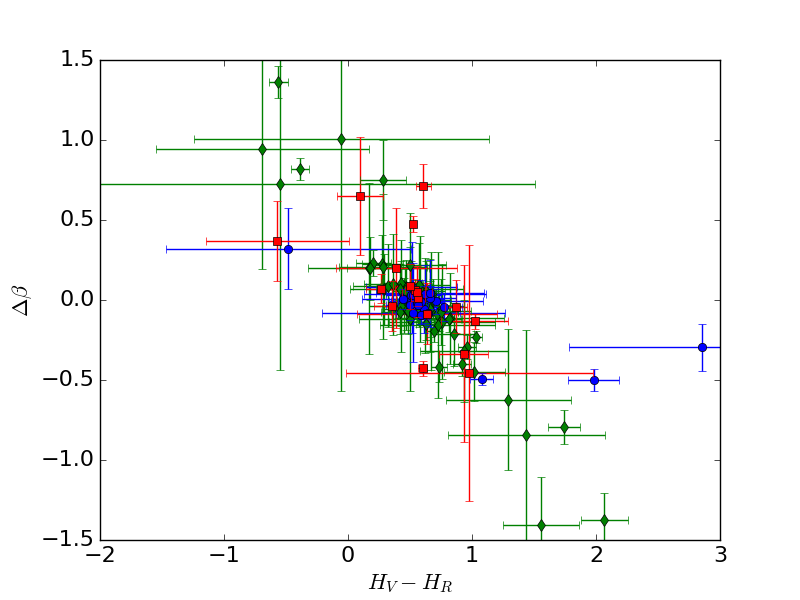}
\includegraphics[scale=0.45]{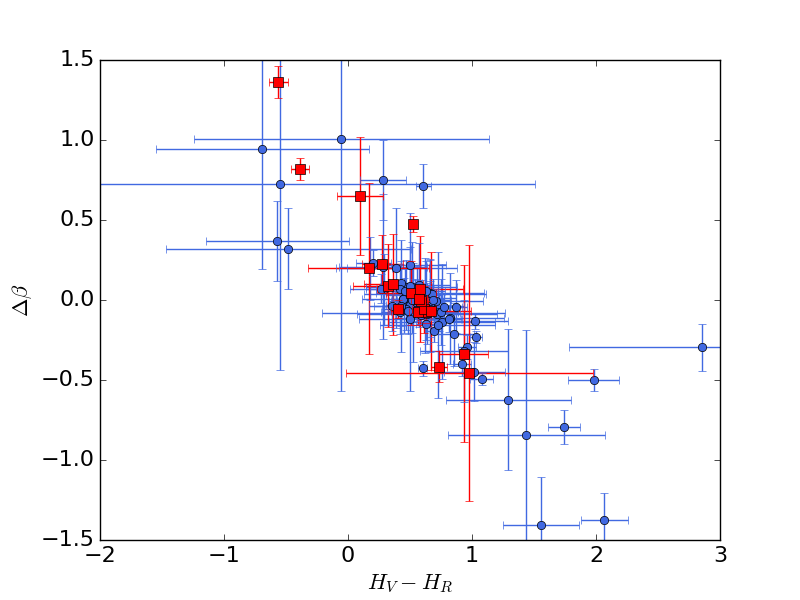}
\caption{Scatter plot of $H_V-H_R$ and $\Delta \beta$. The top panel shows Sample [1] in bins of semi-major axis color-coded as: $a < 40$ AU 
in blue dots, $40 < a < 50 $ AU in green diamonds, $a > 50 $ AU in red squares. The bottom panel shows Sample [2] in blue dots and Sample [3] 
in red squares.}
\label{Fig:colorvsbeta}   
\end{figure}

The most interesting correlation that appears involves $H_V-H_R$ and $\Delta\beta$ (Fig. \ref{Fig:colorvsbeta}). 
The correlation clearly indicates that redder objects have smaller $\Delta\beta$. Physically, this means that, for redder surfaces, 
the phase curves in the $R$ filter are steeper than the ones for the $V$ filter, while the opposite holds for bluer objects. 

\begin{table}
\caption{\small Correlation results. Parameters tested are shown in columns (1, 2), Spearman values (3, 4), and the confidence index CI, in terms of sigma, (5). N is the number of data points. Parameters tested: semi-major axis, $a$, eccentricity $e$, inclination $i$, visible albedo $p_V$, and diameter $D$. The brackets indicates the samples tested (see text).} \label{table:correlations}
\resizebox{!}{0.45\textheight}{\begin{tabular}{|l|l|l|l|l|l|}  
\hline  
                 P1 &            P2 &           $r_s$  &            $P_{r_s}$           &   CI         & N   \\  
\hline
          $H_V$ [1] &           $a$  &        -0.4868  &    $3.96\times^{-8}$ &  (3$\sigma$) & 114 \\   
                    &           $e$ &          0.2380  &    $1.07\times^{-2}$ &  (2$\sigma$) & 114 \\   
                    &           $i$  &        -0.3913  &    $1.66\times^{-5}$ &  (3$\sigma$) & 114 \\   
                    &     $\beta_V$ &         -0.1548  &    $1.01\times^{-1}$ &  (1$\sigma$) & 114 \\   
                    &           $D$ & \textbf{-0.9808} &    $1.69\times^{-46}$&  (3$\sigma$) & 64 \\    
                    &         $p_V$ & \textbf{-0.5807} &    $3.94\times^{-7}$ &  (3$\sigma$) & 64 \\    
     $\beta_V$      &         $p_V$ &          0.0031  &   $9.74\times^{-1}$  &  (<1$\sigma$) & 64 \\   
                    &         $D$   &          0.0215  &   $8.26\times^{-1}$  &  (<1$\sigma$) & 64 \\   
\hline 
          $H_R$ [1] &           $a$ &          -0.4754 &   $1.02\times^{-7}$  &  (3$\sigma$) & 113 \\   
                    &           $e$ &           0.2518 &   $7.12\times^{-3}$  &  (3$\sigma$) & 113 \\   
                    &           $i$ &          -0.3584 &   $9.67\times^{-5}$  &  (3$\sigma$) & 113 \\   
                    &     $\beta_R$ &          -0.1652 &   $8.01\times^{-2}$  &  (1$\sigma$) & 113 \\   
                    &           $D$ & \textbf{-0.9630} &   $1.33\times^{-37}$ &  (3$\sigma$) &  64 \\   
                    &         $p_V$ & \textbf{-0.6389} &   $1.02\times^{-8}$  &  (3$\sigma$) &  64 \\   
     $\beta_R$      &         $p_V$ &          -0.0730 &   $4.59\times^{-1}$  &  (1$\sigma$)  & 64\\   
                    &           $D$ &           0.0098 &   $9.20\times^{-1}$  &  (<1$\sigma$) & 64 \\   
\hline 
$H_V-H_R$ [1]       &$\Delta \beta$   & \textbf{-0.8117} & $8.295\times^{-26}$   & (3$\sigma$)  & 105 \\   
                    &           $a$   &           0.0377 & $7.018\times^{-01}$   &($<1 \sigma$) & 105 \\   
                    &           $e$   &          -0.0054 & $9.558\times^{-01}$   &($<1 \sigma$) & 105 \\   
                    &           $i$   &          -0.1697 & $8.34\times^{-02}$    & (1 $\sigma$) & 105 \\   
                    &          $HV$   &           0.1364 & $1.651\times^{-01}$   & (1 $\sigma$) & 105 \\   
                    &          $HR$   &          -0.0382 & $6.987\times^{-01}$   & ($<1 \sigma$) & 105 \\   
                    &           $D$   &          -0.1016 & $4.204\times^{-01}$   & ($<1 \sigma$) &  64 \\   
                    &         $p_V$   &           0.2224 & $6.21\times^{-02}$    & ($<1 \sigma$) &  64 \\ 
 $\Delta \beta$ [1] &           $a$ &          -0.0074 &  $ 9.397\times^{-01}$   & ($<1 \sigma$) & 105 \\   
                    &           $e$ &           0.0526 &  $ 5.935\times^{-01}$   & ($<1 \sigma$) & 105 \\   
                    &           $i$ &           0.0172 &  $ 8.613\times^{-01}$   & ($<1 \sigma$) & 105 \\   
                    &          $HV$ &          -0.0793 &  $ 4.213\times^{-01}$   & ($<1 \sigma$) & 105 \\   
                    &          $HR$ &           0.0766 &  $ 4.370\times^{-01}$   & ($<1 \sigma$) & 105 \\   
                    &           $D$ &           0.0475 &  $ 7.069\times^{-01}$   & ($<1 \sigma$) &  64 \\   
                    &         $p_V$ &          -0.0892 &  $ 4.796\times^{-01}$   & ($<1 \sigma$) &  64 \\   
\hline 
      $H_V$ [2]     &           $D$ & \textbf{-0.9661} &   $1.01\times^{-28}$  &  (3$\sigma$)  &  47 \\   
                    &         $p_V$ &         -0.4070  &   $4.09\times^{-3}$   &  (3$\sigma$)  &  47 \\   
  $\beta_V$ [2]     &           $D$ &          0.0753  &   $6.1\times^{-01}$    & ($<1 \sigma$)  &  47 \\ 
                    &         $p_V$ &         -0.2096  &   $1.5\times^{-01}$   &  (1$\sigma$)  &  47 \\ 
      $H_R$ [2]     &           $D$ & \textbf{-0.9467} &   $2.81\times^{-24}$  &  (3$\sigma$)  &  47 \\ 
                    &         $p_V$ &         -0.4807  &  $ 5.42\times^{-4}$   &  (3$\sigma$)  &  47 \\ 
 $\beta_R$ [2]      &           $D$ &          0.3261  &   $2.36\times^{-02}$  &  (3$\sigma$)  &  47 \\ 
                    &         $p_V$ &          0.0736  &   $6.18\times^{-01}$  &  (1$\sigma$)  &  47 \\ 
\hline 
$H_V-H_R$ [2]       &$\Delta \beta$ & \textbf{-0.7912} &   $3.34\times^{-19}$ &  (3$\sigma$) &  84 \\   
                    &         $H_V$ &           0.0204 &   $8.53\times^{-1}$  & ($<1 \sigma$)&  84 \\   
                    &         $H_R$ &          -0.1999 &   $6.82\times^{-2}$  &   (1$\sigma$)&  84 \\   
                    &           $D$ &           0.2058 &   $1.60\times^{-1}$  &  (1$\sigma$) &  47 \\   
                    &         $p_V$ &  \textbf{0.5912} &   $9.63\times^{-6}$  &  (3$\sigma$) &  47 \\   
$\Delta \beta$ [2]  & $H_V$         &           0.0710 &  $5.2\times^{-1}$     & ($<1\sigma$) &  84 \\   
                    &         $H_R$ &           0.2652 &  $1.47\times^{-2}$    &  (2$\sigma$) &  84 \\   
                    &           $D$ &          -0.3368 &  $1.92\times^{-2}$    &  (2$\sigma$) &  47 \\   
                    &         $p_V$ &          -0.3769 &  $8.2\times^{-3}$     &  (3$\sigma$) &  47 \\   
\hline 
$H_V$ [3]            &         $D$   &  \textbf{-0.8504}&  1.52$\times^{-05}$  &  (3$\sigma$) &17 \\  
                     &         $p_V$ &  \textbf{-0.6412}&  5.53$\times^{-03}$  &  (3$\sigma$) & 17 \\ 
$\beta_V$            &         $D$   &           0.0948 &  6.826$\times^{-01}$ &  (1$\sigma$) & 17 \\ 
                     &         $p_V$ &           0.0955 &  6.801$\times^{-01}$ &  (1$\sigma$) & 17 \\ 
$H_R$ [3]           &         $D$ &  \textbf{-0.7401} &  6.79$\times^{-04}$ & (3$\sigma$)     & 17 \\ 
                    &         $p_V$   &  \textbf{-0.7862} &  1.82$\times^{-04}$ & (3$\sigma$) & 17 \\
$\beta_R$           &         $D$   &           -0.2583 &  2.580$\times^{-01}$ &  (1$\sigma$) & 17 \\ 
                    &         $p_V$ &           -0.1197 &  6.051$\times^{-01}$ &  (1$\sigma$) & 17 \\ 
\hline 
     $H_V-H_R$ [3]  & $\Delta \beta$ & \textbf{-0.8986} &  3.15$\times^{-08}$  &   (3$\sigma$) &  21 \\    
                    &         $H_V$ &           0.4204 &  5.77$\times^{-02}$  &   (1$\sigma$) &  21 \\    
                    &         $H_R$ &          -0.0045 &  9.84$\times^{-01}$  & ($<1 \sigma$) &  21 \\    
                    &           $D$ &          -0.3872 &  1.24$\times^{-01}$  &   (1$\sigma$) &  17 \\    
                    &         $p_V$ &          -0.3243 &  2.04$\times^{-01}$  &   (1$\sigma$) &  17 \\    
$\Delta \beta$ [3]  &         $H_V$ &          -0.4840 &  2.61$\times^{-02}$ &  (3$\sigma$) &  21 \\   
                    &         $H_R$ &          -0.0708 &  7.603$\times^{-1}$ &($<1 \sigma$) &  21 \\   
                    &         $D$   &           0.3725 &  1.408$\times^{-1}$ &  (1$\sigma$) &  17 \\   
                    &         $p_V$ &           0.3636 &  1.513$\times^{-1}$ &  (1$\sigma$) &  17 \\   
\end{tabular}}
\end{table}

\begin{figure}
\centering
\includegraphics[scale=0.45]{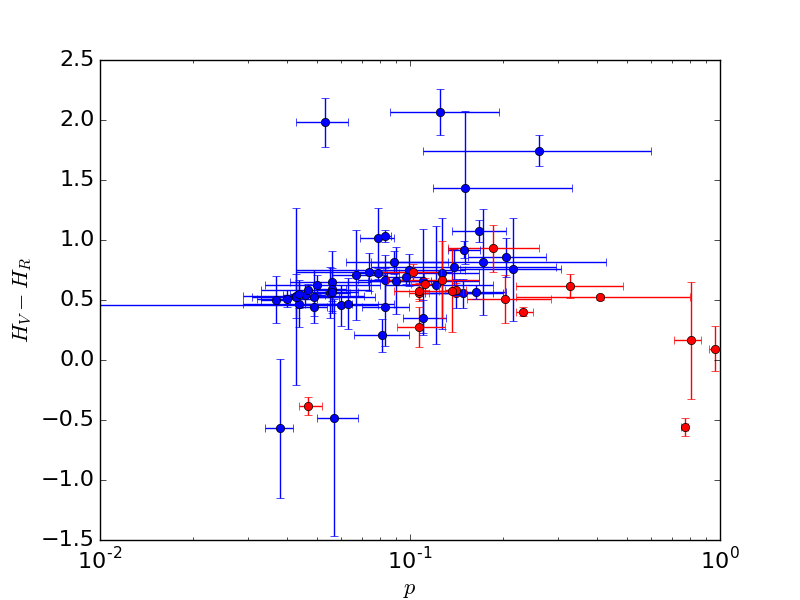}
\caption{Scatter plot of $H_V - H_R $ vs. $p_V$. Sample [2] is shown in blue dots, while Sample [3] is represented with red dots.}
\label{fig:10}
\end{figure}

Another feature of interest, but that has no significance as correlation, between $H_V-H_R$ and $p_V$, can be seen in Fig. \ref{fig:10}. \cite{lacer14}
used a similar scatter plot (but using $S'$ computed from average colors instead of absolute color) to propose that there exist two groups: 
one neutral and dark and other red and bright. Some weak evidence could be seen in our figure as well. We will come back to this issue further ahead.

\noindent
\textbf{Sub-samples by size ([2] and [3]):}\\
\cite{brown12} reported that the surface properties of objects with diameters larger than 500 km are different from smaller objects, for example due to volatile retention \citep[e.g.,][]{schal07}. 
Aiming at understanding if any size-related effect could be seen in our data we separated our objects into ``small'' (Sample [2]) and ``large'' (Sample [3]) according to $H_V=4.5$.

Both sub-samples span more or less the same range in absolute colors and relative phase coefficient, with more extreme values in Sample [2], which seems reasonable due to the lower signal-to-noise ratio of the data.
The most interesting feature is that the anti-correlation between $H_V-H_R$ and $\Delta\beta$ holds for both sub-samples indicating, perhaps, that it is related to surface properties rather than size. Also, as objects from all regions of the trans-Neptunian belt are included (centaurs as well), the correlation is possibly not due to some unknown observational bias (Fig. \ref{Fig:colorvsbeta}).
\begin{figure}
\centering
\includegraphics[scale=0.45]{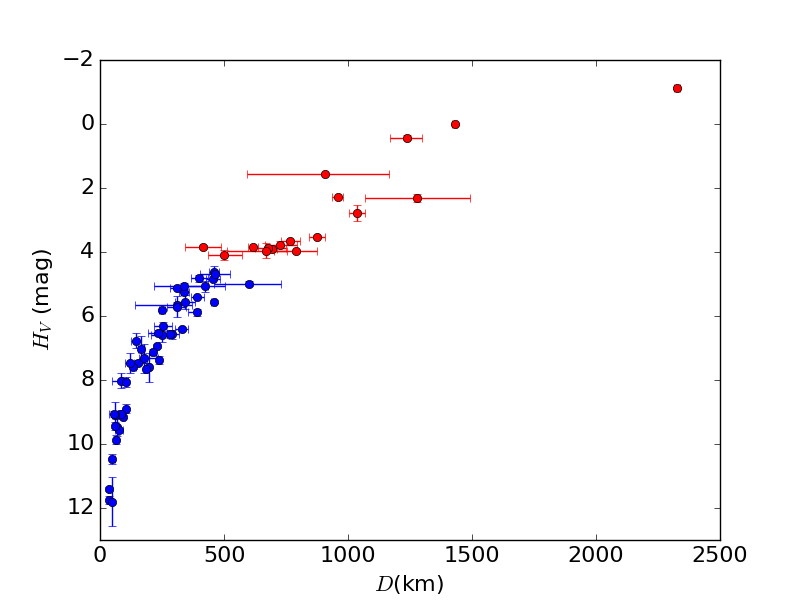}
\includegraphics[scale=0.45]{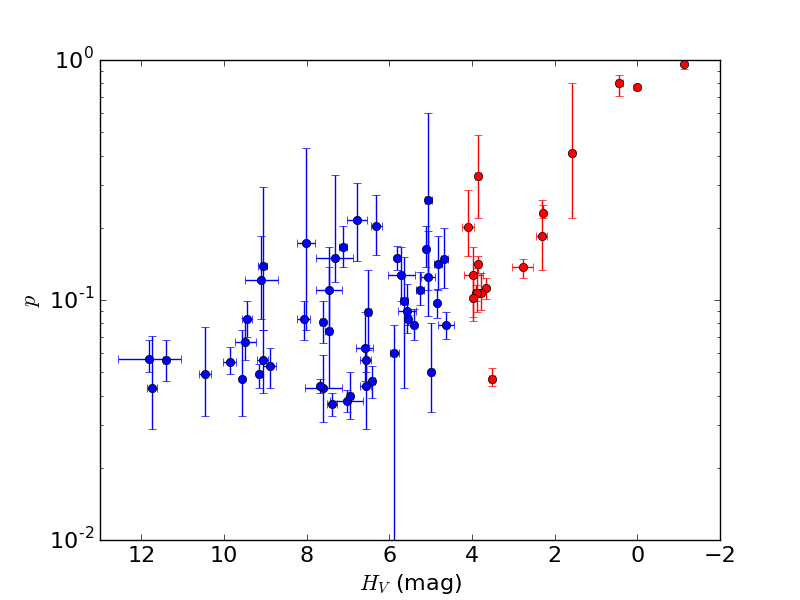}
\caption{Scatter plot of $H_V$ vs. $D$ (top) and $H_V$ vs. $p_V$ (bottom). Samples [2] and [3] are shown in different colors, blue the 
former and red the latter.}
\label{Fig:gapHV_D-p}          
\end{figure}

Sample [2] (Sample [3]) include 84 (21) objects with $V$ and $R$ data, 47 (17) of which have {\it TNOs are cool} diameters and albedos. Sample [2] has on average lower albedos than Sample [3] (Fig.\ref{Fig:gapHV_D-p}, 
bottom panel), which is expected due to the retention of volatiles on the larger objects. 
For Sample [2] the absolute color is correlated with the albedo, implying that high albedo surfaces (with an upper limit in size of 500 km) have redder colors, 
while no similar correlation is seen among Sample [3].

In general, there exists a trend on higher absolute magnitudes $H_V$ (and $H_R$) and albedo.  
We did not use $p_R$, that could have been computed using $H_V-H_R$ and the albedo, as it would not have been a quantity obtained independently from our data.

\section{Discussion and conclusions}\label{sec4}

In this paper we present $H_V$ ($\beta_V$) for 114 objects and $H_R$ ($\beta_R$) for 113. These results were obtained from data observed by ourselves and from the literature. 
Note that 105 objects have both phase curves and therefore absolute colors. In paper 1 we had not included data in the $R$ filter, thus the present work expands our previous results. 
We tested for different correlations, among our data, orbital parameters, and data from the {\it TNOs are cool} survey.

The most interesting correlation we found is between $H_V-H_R$ and $\Delta \beta$. The correlation holds if we consider different bins in semi-major axis (see Fig. \ref{Fig:colorvsbeta}, top) 
and separate between large an small objects (Fig. \ref{Fig:colorvsbeta}, bottom). Therefore, we conclude it is intrinsic to the TNO (and associated) population. 
This correlation indicates that redder objects have steeper phase curves in the $R$ filter than in the $V$ filter, while the opposite is true for bluer objects. 
As many different surfaces types, sizes, and dynamical evolutions are being sampled by our absolute colors we cannot assure that we are seeing an evolutionary effect, 
but probably something related to the porosity and compaction of the surfaces. 
The intrinsic brightness of the object depends of asteroid albedo, which is determined by the surface 
composition, compaction and grains size. There is a dependence of phase coefficient on surface texture \citep{shkur94, shkur94b}.
However, given the inhomogenities of our data base, further studies are needed to clarify this interpretation.

We separated our sample into two sub-groups: large and small, using $H_V=4.5$ as discriminant based on the results of \cite{brown12} in order to use all our dataset. Interestingly, 
there seems to be a gap in this region (Fig. 5 top) which remains to be confirmed. From our search for correlations among the two aforementioned groups (large and small) 
we found that:

\begin{itemize}

\item The correlation between $H_V-H_R$ and $\Delta \beta$ holds for both of them. As we do not expect the surface composition to be the same in both groups 
 (in fact it is clear that the large objects do not have objects as red ad the small population) and the albedo distributions are different, 
 we consider that the correlation is due to surface micro-structure (compaction, grain size) in a yet-to-be-understood way. 
 Surface temperature does not seem to be a key factor here either, 
 at least in first approximation, as the correlation holds for different bins in semi-major axis as well.
 
\item There exists a significant correlation between $H_V-H_R$ and $p_V$ for the small objects (upper size limit at $\sim 500$ km). 
According to this correlation the redder objects have higher albedos. For the larger objects there is a marginal (opposite) trend. 

\item Regarding the small objects, the high-albedo tend, as well, to be the larger objects as shown in Fig. \ref{Fig:gapHV_D-p}. According to our understanding 
of collisional and resurfacing models, an initially neutral bright surface (ice-covered) under irradiation will decrease its albedo in the visible faster 
than in the red. Further irradiation will decrease the albedo in the red letting a dark carbon-covered surface  with a neutral 
slope \citep[e.g.,][and references]{strazz91,thompson87,hud08}. 
The main opposing mechanism is collisional, that craterize the surface exposing sub-superficial material \cite[see][]{hutto02}. 
Therefore, the larger ``small'' objects, of higher albedo and red, are probably well processed, but not yet in the last stages of irradiation, 
while the smaller, bluer and lower-albedo, could be objects that are very processed. This is very curious, because these smaller objects (Fig. 6-right) should have suffered more impacts than larger ones \citep[e.g.,][]{Dohn71, farinelladavis92} and, in principle, there should be at least some objects with higher albedos.

\end{itemize}

We also searched for correlations among the different dynamical classes: Centaurs, Classical TNOs, 
Scattering objects (including Detached objects, and Resonant objects, following \cite{gladman08}. 
Although with a lower number of objects, the correlation between $H_V - H_R$ and $\Delta\beta$ appears 
in all subpopulations, pointing even more towards a property shared by all these minor bodies and that deserves further analysis.

Following part of the discussion drawn in paper 1, it becomes clear that, although the $\beta$ distribution are clearly unimodal and 
that about 60 \% of the objects are close to the mode of the distribution, a non-negligible fraction of objects have values that can 
differ by a significant amount of the mode. Therefore an ``average'' value of the phase coefficient must be taken, and used, with extreme caution.

A phase curve with negative values of $\beta$ do not have a direct physical interpretation in terms of photometric models, see our discussion in paper 1. 
Nevertheless, some plausible explanations are (i) underestimation of rotational amplitude, which could account for values of reduced magnitudes different 
than expected by our simple model (Eq. \ref{eq_rand}),
(ii) the presence of material surrounding the body 
(ring systems, satellites, or binaries) which modify the total reflecting area as seen from Earth, and (iii) faint cometary-like activity. Further deeper, in quality 
and quantity, photometric studies are needed in order to discern between these scenarios.

The absolute magnitudes and phase coefficients have been obtained from heterogeneous sources, with a variety of precisions, from a wide distribution 
of telescopes, instruments, and filters. Nevertheless, we have used homogeneous techniques to analyze them and produce an accurate database, although 
probably not as precise as desired, especially due to the large uncertainties introduced by unknown rotational properties. Also, our results are, in 
a way, mean values, as observations covering large intervals of time are being used here and some objects are known to suffer changes in relatively short 
time-scales, for instance the ring bearer Chariklo. Nevertheless, the statistical significance of our database is robust and we intend to continue 
increasing it, including new observations reported in the literature and our own forthcoming observations.

\section*{Acknowledgments}
We are very grateful to the anonymous referee for valuable comments which helped us to improve this paper.
Based in part on observations collected at the German-Spanish Astronomical Center, Calar Alto, operated jointly by Max-Planck- Institut f\"ur Astronomie 
and Instituto de Astrof\'isica de Andaluc\'ia (CSIC). Partially based on observations obtained at the Southern Astrophysical Research (SOAR) telescope, 
which is a joint project of the Minist\'erio da Ci\^encia, Tecnologia, e Inova\c c\~ao (MCTI) da Rep\'ublica Federativa do Brasil, the U.S. National Optical 
Astronomy Observatory (NOAO), the University of North Carolina at Chapel Hill (UNC), and Michigan State University (MSU).
CAL acknowledges support from CNPq.
AAC acknowledges support from CNPq and FAPERJ.
Part of the research leading to these results has received funding from the European Union's Horizon 2020 Research and Innovation Programme, 
under Grant Agreement No. 687378. PSS and JLO would like to acknowledge financial support by the Spanish grant AYA-2014-56637-C2-1-P and 
the Proyecto de Excelencia de la Junta de Andaluc\'ia J.A. 2012-FQM1776.
\begin{table*}
	\caption{Absolute Magnitudes}\label{table:magnitudesH}
	\begin{tabular}{|l|l|l|l|l|l|l|l|l|}  
\hline
                   Object           &        $H_V$      &     $\beta_V$            &  $N_V$  &      $H_R$        &        $\beta_R$      &  $N_R$    &$\Delta m$  &         Ref. \\ 
                                    &        (mag)      & (mag per degree)         &         &      (mag)        &    (mag per degree)   &           &            &             \\  
\hline
         15760   1992 QB$_{1}$     & $  7.83\pm0.05 $  &   $ -0.193\pm0.05 $       &    3    &  $ 6.87\pm0.05 $  &   $ 0.102\pm0.05 $    &     3     &    0.04   &                              Al16 \\
         15788    1993  SB          & $  7.99\pm0.18 $  &   $  0.373\pm0.13 $      &    5    &  $ 7.70\pm0.20 $  &   $ 0.307\pm0.17 $    &     4    &    0.14    &                              Al16 \\
         15789    1993  SC          & $  7.39\pm0.02 $  &   $  0.050\pm0.01 $      &    8    &  $ 6.71\pm0.02 $  &   $ 0.015\pm0.02 $    &     6     &    0.04   &                              Al16 \\
                  1994  EV$_{33}$   & $  8.18\pm0.12 $  &   $ -0.801\pm0.19 $      &    3    &  $ 7.75\pm0.11 $  &   $-0.907\pm0.18 $    &     3     &    0.14    &                              Al16 \\
         16684    1994  JQ$_{1}$    & $  7.03\pm0.14 $  &   $  0.568\pm0.18 $      &    5    &  $ 6.75\pm0.11 $  &   $-0.181\pm0.17 $    &     5     &    0.14    &                              Al16 \\
                  1994  TB          & $  8.01\pm0.22 $  &   $  0.132\pm0.15 $      &    9    &  $ 7.19\pm0.37 $  &   $ 0.247\pm0.23 $    &     7     &    0.34   &                              Al16 \\
                  1994  VK8         & $  7.83\pm0.91 $  &   $ -0.174\pm0.97 $      &    3    &  $   \cdots    $  &   $    \cdots    $    &  $\cdots$ &    0.42   &                              Al16 \\
                  1995  HM5         & $  8.31\pm0.08 $  &   $  0.037\pm0.08 $      &    5    &  $ 7.89\pm0.17 $  &   $-0.032\pm0.12 $    &     4     &    0.14    &                             Al16 \\
                  1995  QY$_{9}$    & $  8.13\pm0.53 $  &   $ -0.110\pm0.53 $      &    3    &  $  \cdots     $  &   $    \cdots    $    &  $\cdots$ &    0.60   &                              Al16 \\
         24835    1995  SM$_{55}$   & $  4.58\pm0.17 $  &   $  0.137\pm0.19 $      &   7    &  $ 4.15\pm0.18 $  &   $ 0.177\pm0.20 $    &      7    &    0.19   &                              Al16 \\
         26181    1996  GQ$_{21}$   & $  5.07\pm0.04 $  &   $  0.857\pm0.12 $      &    6    &  $ 4.46\pm0.02 $  &   $ 0.144\pm0.05 $    &      9    &    0.14    &                 Al16, \textbf{TW} \\
                  1996  RQ$_{20}$   & $  6.97\pm0.08 $  &   $ 0.56\pm0.10 $        &    4    &  $ 6.67\pm0.29 $  &   $ 0.269\pm0.26 $      &      4    &    0.14    &                              Al16 \\
                  1996  RR$_{20}$   & $  6.98\pm0.18 $  &   $  0.391\pm0.19 $      &    3    &  $ 6.40\pm0.17 $  &   $ 0.297\pm0.18 $    &      3    &    0.14    &                              Al16 \\
         19299    1996  SZ$_{ 4}$   & $  8.56\pm0.08 $  &   $  0.306\pm0.07 $      &    4   &  $ 8.04\pm0.08 $  &   $ 0.308\pm0.07 $    &      3    &    0.14    &                              Al16 \\
                  1996  TK$_{66}$   & $  7.02\pm0.20 $  &   $ -0.278\pm0.22 $      &    3    &  $ 6.52\pm0.20 $  &   $-0.499\pm0.22 $    &      3    &    0.14    &                              Al16 \\
         15874    1996  TL$_{66}$   & $  5.25\pm0.09 $  &   $  0.375\pm0.11 $      &    5    &  $ 4.90\pm0.10 $  &   $ 0.414\pm0.10 $    &      3    &    0.12   &                              Al16 \\
         19308    1996  TO$_{66}$   & $  4.81\pm0.19 $  &   $  0.174\pm0.23 $      &   10    &  $ 4.31\pm0.36 $  &   $ 0.291\pm0.38 $    &      8    &    0.14    &      JL01, Se02,Al16,\textbf{TW} \\
         15875    1996  TP$_{66}$   & $  7.46\pm0.08 $  &   $  0.126\pm0.07 $      &    5    &  $ 6.73\pm0.13 $  &   $ 0.199\pm0.09 $    &      3    &    0.14    &                             Al16 \\
        118228    1996  TQ$_{66}$   & $  8.00\pm0.42 $  &   $ -0.414\pm0.67 $      &    4    &  $    \cdots   $  &   $    \cdots    $    &     $\cdots$    &    0.14    &                             Al16 \\
                  1996  TS$_{66}$   &  $  6.53\pm0.16$  &   $ 0.084\pm0.21   $    &     4    &  $  \cdots   $    &   $ \cdots$           & $\cdots$      &0.14 &Da00,JL01,JL98,RT99 \\
         33001    1997  CU$_{29}$   & $ 6.80\pm0.07  $  &   $  0.075\pm0.12 $      &    4    &  $ 6.15\pm0.07 $  &   $ 0.133\pm0.12 $    &      4    &    0.14    &                             Al16 \\
                  1997   QH$_{4}$   & $ 7.21\pm0.25  $  &   $  0.450\pm0.22 $      &    4    &  $ 6.27\pm0.25 $  &   $ 0.766\pm0.22 $    &      4    &    0.14    &                             Al16 \\
         24952    1997   QJ$_{4}$   & $ 7.75\pm0.11  $  &   $  0.290\pm0.10 $      &    5    &  $ 7.26\pm0.11 $  &   $ 0.357\pm0.10 $    &      4    &    0.14    &                             Al16 \\
         33128    1998  BU$_{48}$   & $ 5.71\pm2.00  $  &   $  0.967\pm1.12 $      &    4    &  $ 6.26\pm0.47 $  &   $ 0.244\pm0.26 $    &     12    &    0.68   &                 Al16, \textbf{TW} \\
         91133    1998 HK$_{151}$   & $  7.33\pm0.05 $  &   $  0.127\pm0.08 $      &    5    &  $ 6.87\pm0.07 $  &   $ 0.071\pm0.10 $    &      5    &    0.14    &                              Al16 \\
        385194    1998  KG$_{62}$   & $  7.64\pm0.10 $  &   $ -0.747\pm0.13 $      &    3    &  $ 7.01\pm0.10 $  &   $-0.803\pm0.13 $    &      3    &    0.14    &                              Al16 \\
         85633    1998  KR$_{65}$   & $    \cdots    $  &   $   \cdots      $      &$\cdots$ &  $-1.91\pm1.43 $  &   $ 7.187\pm1.21 $    &      3    &    0.14    &                         Bo02,TR03 \\
         26308    1998 SM$_{165}$   & $  5.93\pm0.36 $  &   $  0.448\pm0.37 $      &    3    &  $    \cdots   $  &   $     \cdots   $    & $\cdots$  &    0.14    &                              Al16 \\
         35671    1998 SN$_{165}$   & $  5.87\pm0.10 $  &   $ -0.031\pm0.11 $      &    6    &  $ 5.42\pm0.12 $  &   $ 0.005\pm0.12 $    &      5    &    0.14    &                              Al16 \\
                  1998  UR$_{43}$   & $  9.04\pm0.11 $  &   $ -0.763\pm0.20 $      &    3    &  $ 8.42\pm0.11 $  &   $-0.728\pm0.21 $    &      3    &    0.14    &                             Al16 \\ 
         33340    1998  VG$_{44}$   & $  6.60\pm0.20 $  &   $  0.226\pm0.15 $      &    3    &  $ 6.13\pm0.05 $  &   $ 0.145\pm0.05 $    &      5    &    0.10   &          Se02, Al16, \textbf{TW} \\ 
                  1999 CD$_{158}$   & $  5.28\pm0.23 $  &   $  0.092\pm0.30 $      &    3    &  $ 5.00\pm0.26 $  &   $-0.114\pm0.33 $    &      3    &    0.14    &                             Al16 \\ 
         26375    1999   DE$_{9}$   & $  5.11\pm0.02 $  &   $  0.182\pm0.03 $      &   36    &  $ 4.55\pm0.05 $  &   $ 0.167\pm0.04 $    &     11    &    0.10   &          Se02, Al16, \textbf{TW} \\ 
                  1999  HS$_{11}$   & $  6.84\pm0.86 $  &   $  0.227\pm1.12 $      &    3    &  $ 6.90\pm0.81 $  &   $-0.779\pm1.10 $    &      3    &    0.14    &                             Al16 \\ 
         40314    1999  KR$_{16}$   & $  6.31\pm0.13 $  &   $ -0.124\pm0.18 $      &    4    &  $ 5.46\pm0.07 $  &   $ 0.091\pm0.07 $    &     15    &    0.18   &                Al16, \textbf{TW} \\ 
         44594    1999   OX$_{3}$   & $  7.60\pm0.06 $  &   $  0.108\pm0.04 $      &   14    &  $ 7.40\pm0.12 $  &   $-0.122\pm0.07 $    &     19    &    0.11   &    BA03, Th12, Al16, \textbf{TW} \\ 
         86047    1999   OY$_{3}$   & $  6.44\pm0.13 $  &   $  0.272\pm0.15 $      &    3    &  $ 6.26\pm0.12 $  &   $ 0.075\pm0.11 $    &      5    &    0.14    &                              Al16 \\ 
                  1999 RY$_{215}$   & $    \cdots    $  &   $    \cdots   $        &$\cdots$ &  $ 6.60\pm0.10 $  &   $ 0.429\pm0.13 $    &      3    &    0.14    &                    Bo02,Do01,Sn10 \\ 
         47171    1999  TC$_{36}$   & $  5.39\pm0.02 $  &   $  0.110\pm0.02 $      &   45    &  $ 4.67\pm0.03 $  &   $ 0.195\pm0.03 $    &      7    &    0.070   &                               Al16 \\ 
         29981    1999  TD$_{10}$   & $  9.09\pm0.38 $  &   $  0.036\pm0.11 $      &   27    &  $ 8.45\pm0.40 $  &   $ 0.122\pm0.15 $    &     16    &    0.650   &                   Ro03, Mu04, Al16 \\ 
        121725    1999 XX$_{143}$   & $  9.09\pm0.27 $  &   $  0.066\pm0.20 $      &    4    &  $ 8.57\pm0.24 $  &   $-0.012\pm0.19 $    &      4    &    0.14    &            BA03, Al16, \textbf{TW} \\ 
         47932    2000 GN$_{171}$   & $  6.77\pm0.24 $  &   $ -0.100\pm0.18 $      &   30    &  $ 6.01\pm0.35 $  &   $ 0.035\pm0.30 $    &     13    &    0.61   &      SJ02, Ca12, Al16, \textbf{TW} \\ 
        138537    2000  OK$_{67}$   & $  6.63\pm0.86 $  &   $  0.087\pm0.65 $      &    3    &  $    \cdots   $  &   $     \cdots   $    &      0    &    0.14    &                               Al16 \\ 
         82075    2000 YW$_{134}$   & $  4.38\pm0.68 $  &   $  0.373\pm0.55 $      &    3    &  $ 3.40\pm0.72 $  &   $ 0.832\pm0.58 $    &      3    &    0.10   &                               Al16 \\ 
         63252    2001  BL$_{41}$   & $ 11.74\pm0.12 $  &   $  0.027\pm0.03 $      &    4    &  $11.21\pm0.13 $  &   $ 0.033\pm0.03 $    &      4    &    0.14    &                   Al16, \textbf{TW} \\ 
        150642    2001  CZ$_{31}$   & $    \cdots    $  &   $    \cdots     $      &$\cdots$ &  $ 5.54\pm0.14 $  &   $ 0.111\pm0.16 $    &      3    &    0.21   &                               SJ02 \\ 
         82158    2001 FP$_{185}$   & $  6.40\pm0.06 $  &   $  0.140\pm0.04 $      &   6     &  $ 5.87\pm0.05 $  &   $ 0.078\pm0.04 $    &     7     &    0.06   &                  Al16,\textbf{TW} \\ 
         82155    2001 FZ$_{173}$   & $  6.12\pm0.08 $  &   $  0.339\pm0.08 $      &   4     &  $ 5.62\pm0.09 $  &   $ 0.253\pm0.10 $    &     6     &    0.06   &           SJ02, Al16, \textbf{TW} \\ 
                  2001 KA$_{77}$    & $  5.64\pm0.09 $  &   $  0.130\pm0.11 $      &   3     &  $ 4.89\pm0.09 $  &   $ 0.206\pm0.16 $    &     3     &    0.14    &                              Al16 \\ 
                  2001 KD$_{77}$    & $  6.52\pm0.07 $  &   $ -0.005\pm0.06 $      &   4     &  $ 5.71\pm0.06 $  &   $ 0.111\pm0.05 $    &     4     &    0.07  &                              Al16 \\ 
                  2001 QC$_{298}$   & $     \cdots   $  &   $      \cdots   $      &$\cdots$ &  $ 6.06\pm0.10 $  &   $ 0.331\pm 0.1 $    &     3     &    0.14    &              Sn10,SS09,\textbf{TW} \\ 
                  2001 QY$_{297}$   & $     \cdots   $  &   $      \cdots   $      &$\cdots$ &  $ 5.50\pm0.24 $  &   $-0.295\pm 0.2 $    &     7     &    0.49   &                 Th12, \textbf{TW} \\ 
         42301    2001 UR$_{163}$   & $  4.52\pm0.06 $  &   $  0.363\pm0.11 $      &   3     &  $ 3.65\pm0.06 $  &   $ 0.404\pm0.11 $    &     3     &    0.08    &                              Al16 \\ 
         55565    2002 AW$_{197}$   & $  3.65\pm0.02 $  &   $  0.077\pm0.03 $      &   38    &  $ 3.02\pm0.04 $  &   $ 0.151\pm0.06 $    &     4     &    0.04    &                              Al16 \\ 
                  2002 GP$_{32}$    & $  7.13\pm0.02 $  &   $ -0.134\pm0.03 $      &    4    &  $ 6.53\pm0.02 $  &   $ 0.292\pm0.02 $    &     4     &    0.03   &                              Al16 \\ 
         95626    2002 GZ$_{32}$    & $  7.38\pm0.12 $  &   $  0.072\pm0.05 $      &   30    &  $ 6.88\pm0.15 $  &   $ 0.106\pm0.06 $    &     5     &    0.15   &                              Al16 \\ 
                  2002 KW$_{14}$    & $    \cdots    $  &   $    \cdots     $      &$\cdots$ &  $ 6.34\pm0.40 $  &   $-1.420\pm0.37 $    &     5     &    0.25   &                 Th12, \textbf{TW} \\ 
         119951    2002 KX$_{14}$   & $  4.83\pm0.03 $  &   $  0.277\pm0.03 $      &   21    &  $ 4.14\pm0.04 $  &   $ 0.468\pm0.06 $    &     5     &     0.05   &              Re13                 Al16 \\ 
         250112    2002 KY$_{14}$   & $ 11.80\pm0.76 $  &   $ -0.273\pm0.19 $      &    4    &  $12.28\pm0.62 $  &   $-0.593\pm0.16 $    &     4     &    0.13   &                              Al16 \\ 
                   2002 PN$_{34}$   & $  8.61\pm0.05 $  &   $  0.089\pm0.02 $      &   57    &  $    \cdots   $  &   $    \cdots    $    & $\cdots$  &    0.18   &                              Al16 \\ 
         \hline
	\end{tabular}                                        
\end{table*}

\addtocounter{table}{-1}
\begin{table*}
	\caption{Absolute Magnitudes}
	\begin{tabular}{|l|l|l|l|l|l|l|l|l|}  
\hline
                   Object           &        $H_V$      &     $\beta_V$            &  $N_V$  &     $H_R$         &        $\beta_R$      &  $N_R$    &$\Delta m$ &         Ref. \\ 
                                    &        (mag)      & (mag per degree)         &         &      (mag)        &    (mag per degree)   &           &           &             \\  
\hline
          55637    2002 UX$_{25}$   & $  3.90\pm0.04 $  &   $  0.104\pm0.05 $      &  46     &  $ 3.34\pm0.05 $  &   $ 0.176\pm0.06 $    &   17      &    0.21  &                  Al16,\textbf{TW} \\ 
          55638    2002 VE$_{95}$   & $  5.81\pm0.03 $  &   $  0.088\pm0.02 $      &  43     &  $ 4.89\pm0.06 $  &   $ 0.487\pm0.07 $    &    4      &    0.08  &                              Al16 \\ 
         127546    2002 XU$_{93}$   & $  7.03\pm0.40 $  &   $  0.496\pm0.17 $      &   5     &  $ 7.60\pm0.41 $  &   $ 0.125\pm0.17 $    &    5      &    0.14   &                              Al16 \\ 
         208996    2003 AZ$_{84}$   & $  3.77\pm0.11 $  &   $  0.074\pm0.11 $      &   5     &  $ 3.49\pm0.11 $  &   $-0.151\pm0.13 $    &    5      &    0.14  &                              Al16 \\ 
         120061    2003 CO$_{1}$    & $  9.14\pm0.05 $  &   $  0.092\pm0.01 $      &   5     &  $ 8.70\pm0.05 $  &   $ 0.084\pm0.01 $    &    5      &    0.07  &                              Al16 \\ 
         133067    2003 FB$_{128}$  & $  6.92\pm0.60 $  &   $  0.422\pm0.53 $      &   3     &  $ 7.61\pm0.60 $  &   $-0.519\pm0.52 $    &    3      &    0.14   &                              Al16 \\ 
                   2003 FE$_{128}$  & $  7.38\pm0.34 $  &   $ -0.348\pm0.29 $      &   5     &  $ 6.08\pm0.36 $  &   $ 0.274\pm0.32 $    &    5      &    0.14   &                              Al16 \\ 
         120132    2003 FY$_{128}$  & $  4.63\pm0.18 $  &   $  0.534\pm0.14 $      &   7     &  $ 3.61\pm0.16 $  &   $ 0.983\pm0.11 $    &    6      &    0.15  &                              Al16 \\ 
         385437    2003 GH$_{55}$   & $  7.31\pm0.44 $  &   $ -0.878\pm0.46 $      &   3     &  $ 5.88\pm0.44 $  &   $-0.034\pm0.46 $    &    3      &    0.14   &                              Al16 \\ 
         120178    2003 OP$_{32}$   & $  4.05\pm0.21 $  &   $  0.056\pm0.19 $      &  10     &  $ 3.73\pm0.17 $  &   $-0.033\pm0.16 $    &    9      &    0.18   &                              Al16 \\ 
                   2003 QW$_{90}$   & $  6.35\pm0.45 $  &   $ -1.137\pm0.51 $      &   3     &  $    \cdots   $  &   $    \cdots    $    & $\cdots$  &    0.14   &                             Al16 \\ 
                   2003 UY$_{117}$  & $    \cdots    $  &    $    \cdots    $      &$\cdots$ &  $ 5.60\pm0.10 $  &   $ 0.280\pm0.11 $    &    3      &    0.14   &                             Al16 \\ 
         416400    2003 UZ$_{117}$  & $  5.23\pm0.10 $  &   $  0.133\pm0.10 $      &    5    &  $ 4.81\pm0.11 $  &   $ 0.209\pm0.09 $    &    4      &    0.14   &                             Al16 \\ 
                   2003 UZ$_{413}$  & $  4.36\pm0.17 $  &   $  0.143\pm0.22 $      &    3    &  $ 3.99\pm0.16 $  &   $ 0.044\pm0.22 $    &    3      &    0.14   &                             Al16 \\ 
         136204    2003 WL$_{7}$    & $  8.89\pm0.14 $  &   $  0.088\pm0.04 $      &    4    &  $ 6.91\pm0.14 $  &   $ 0.588\pm0.04 $    &    3      &    0.05  &                              Al16 \\ 
         120216    2004 EW$_{95}$   & $  6.57\pm0.13 $  &   $  0.080\pm0.09 $      &    5    &  $ 6.10\pm0.13 $  &   $ 0.135\pm0.09 $    &    5      &    0.14   &                               Al16 \\ 
         307982    2004 PG$_{115}$  & $  4.95\pm0.45 $  &   $  0.445\pm0.34 $      &    8    &  $ 4.56\pm0.17 $  &   $ 0.243\pm0.14 $    &    9      &    0.14   &                 Al16, \textbf{TW} \\ 
                   2004 PT$_{107}$  & $    \cdots    $  &   $    \cdots     $      &$\cdots$ &  $ 6.33\pm1.02 $  &   $-0.347\pm0.76 $    &    3      &    0.14   &                              Al16 \\ 
                   2004 TY$_{364}$  & $  4.51\pm0.13 $  &   $  0.145\pm0.10 $      &   32    &  $    \cdots   $  &   $    \cdots    $    & $\cdots$  &    0.22  &                               Al16 \\ 
         144897    2004 UX$_{10 }$  & $  4.82\pm0.09 $  &   $  0.060\pm0.10 $      &    8    &  $ 4.26\pm0.07 $  &   $ 0.062\pm0.07 $    &    8      &    0.08  &                               Al16 \\ 
         230965    2004 XA$_{192}$  & $  5.05\pm0.08 $  &   $ -0.174\pm0.07 $      &    5    &  $ 3.31\pm0.09 $  &   $ 0.620\pm0.07 $    &    6      &    0.07  &                  Al16, \textbf{TW} \\ 
                   2005 GE$_{187}$  & $    \cdots    $  &   $    \cdots     $      & $\cdots$&  $ 7.13\pm0.18 $  &   $ 0.065\pm0.13 $    &    3      &    0.14   &                   Ca12,\textbf{TW} \\ 
                   2005 QU$_{182}$  & $  3.85\pm0.06 $  &   $  0.277\pm0.10 $      &     5   &  $ 3.23\pm0.06 $  &   $ 0.336\pm0.10 $    &    5      &    0.12   &                               Al16 \\ 
                   2005 RM$_{43}$   & $  4.70\pm0.08 $  &   $ -0.027\pm0.06 $      &     6   &  $ 4.44\pm0.08 $  &   $-0.098\pm0.06 $    &    5      &    0.04  &                               Al16 \\ 
                   2005 RN$_{43}$   & $  3.88\pm0.05 $  &   $  0.139\pm0.04 $      &    11   &  $ 3.30\pm0.03 $  &   $ 0.133\pm0.03 $    &   10      &    0.06  &            Re13, Al16, \textbf{TW} \\ 
                   2005 RR$_{43}$   & $  4.25\pm0.06 $  &   $ -0.003\pm0.06 $      &     5   &  $ 3.75\pm0.06 $  &   $ 0.160\pm0.06 $    &    4      &    0.06  &            Re13,             Al16 \\ 
                   2005 TB$_{190}$  & $  4.67\pm0.08 $  &   $  0.051\pm0.10 $      &     8   &  $ 4.12\pm0.08 $  &   $-0.001\pm0.11 $    &   12      &    0.12  &                  Al16, \textbf{TW} \\ 
                   2005 UQ$_{513}$  & $  4.09\pm0.14 $  &   $ -0.130\pm0.14 $      &     3   &  $ 3.58\pm0.13 $  &   $-0.174\pm0.13 $    &    4      &    0.06  &             Sn10,Al16*,\textbf{TW} \\ 
                   2007 OC$_{10}$   & $  5.70\pm0.32 $  &   $ -0.115\pm0.32 $      &     4   &  $ 4.98\pm0.32 $  &   $ 0.042\pm0.32 $    &    4      &    0.14   &                  Al16, \textbf{TW} \\ 
                   2007 OR$_{10}$   & $  2.31\pm0.13 $  &   $  0.255\pm0.34 $      &     7   &  $ 1.38\pm0.14 $  &   $ 0.590\pm0.43 $    &    4      &    0.09   &                               Al16 \\ 
                   2008 FC$_{76}$   & $  9.48\pm0.26 $  &   $  0.101\pm0.05 $      &     3   &  $ 8.77\pm0.26 $  &   $ 0.110\pm0.05 $    &    4      &    0.14   &                                Al16 \\ 
                   2008 OG$_{19}$   & $    \cdots    $  &   $    \cdots    $       &$\cdots$ &  $ 6.46\pm0.19 $  &   $-1.787\pm0.17 $    &    4      &    0.14   &                         \textbf{TW} \\ 
                   2013 AZ$_{60}$    & $  10.4\pm0.19 $  &   $  0.030\pm0.03 $      &     3   &  $ 9.43\pm0.19 $  &   $ 0.164\pm0.03 $    &    3      &    0.14   &                                Al16 \\ 
              55576    Amycus       & $  8.07\pm0.16 $  &   $  0.127\pm0.05 $      &     5   &  $ 7.41\pm0.13 $  &   $ 0.113\pm0.04 $    &    5      &    0.16  &              BA03,Al16,\textbf{TW} \\ 
              8405    Asbolus       & $  9.06\pm0.13 $  &   $  0.072\pm0.03 $      &    43   &  $ 8.41\pm0.21 $  &   $ 0.155\pm0.06 $    &    6      &    0.55  &              BA03,Al16,\textbf{TW} \\ 
              54598    Bienor       & $  7.59\pm0.45 $  &   $  0.188\pm0.19 $      &    59   &  $ 7.06\pm0.58 $  &   $ 0.267\pm0.24 $    &    5      &    0.75  &              BA03,Al16,\textbf{TW} \\ 
                       Borasisi     & $  6.03\pm0.03  $  &   $  0.23\pm0.06$       &    3    &  $ \cdots $  &   $ \cdots  $    &   $ \cdots  $       &    0.05  &              BA03,Al16,\textbf{TW} \\               
              65489      Ceto       & $  6.57\pm0.12 $  &   $  0.195\pm0.09 $      &    9    &  $ 5.98\pm0.12 $  &   $ 0.209\pm0.09 $    &   10      &    0.13  &                   Al16,\textbf{TW} \\ 
              19521     Chaos       & $  4.98\pm0.06 $  &   $  0.102\pm0.07 $      &     6   &  $ 4.36\pm0.04 $  &   $ 0.254\pm0.06 $    &    7      &    0.10  &              SJ02,Al16,\textbf{TW} \\ 
              10199  Chariklo       & $  6.94\pm0.05 $  &   $  0.049\pm0.01 $      &    21   &  $ 6.42\pm0.04 $  &   $ 0.021\pm0.01 $    &   35      &    0.10  &         Ga16,BA03,Al16,\textbf{TW} \\ 
              2060     Chiron       & $  7.11\pm0.08 $  &   $ -0.410\pm0.03 $      &    8    &  $ 6.04\pm0.02 $  &   $ 0.080\pm0.00 $    &    54      &    0.09  &         Ga16,BA03,Al16,\textbf{TW} \\ 
              83982   Crantor       & $  9.09\pm0.40 $  &   $  0.109\pm0.14 $      &     6   &  $ 8.47\pm0.28 $  &   $ 0.074\pm0.10 $    &    5      &    0.34  &              BA03,Al16,\textbf{TW} \\ 
              52975  Cyllarus       & $  9.06\pm0.10 $  &   $  0.171\pm0.06 $      &     6   &  $ 8.29\pm0.10 $  &   $ 0.218\pm0.06 $    &    6      &    0.14   &               BA03,Al16,\textbf{TW} \\ 
             60558   Echeclus       & $  9.86\pm0.14 $  &   $  0.056\pm0.05 $      &    11   &  $ 9.30\pm0.14 $  &   $ 0.076\pm0.05 $    &   13      &    0.24  &              BA03,Ro05,\textbf{TW} \\ 
             31824     Elatus       & $ 10.46\pm0.14 $  &   $  0.088\pm0.02 $      &    13   &  $ 9.93\pm0.16 $  &   $ 0.059\pm0.03 $    &   16      &    0.24  &              BA02,Al16,\textbf{TW} \\ 
             136199      Eris       & $ -1.12\pm0.02 $  &   $  0.135\pm0.05 $      &    76   &  $-1.22\pm0.18 $  &   $-0.516\pm0.36 $    &    9      &    0.10  &    DM09,Ra07,Ca06,Al16,\textbf{TW} \\ 
            136108     Haumea       & $  0.43\pm0.07 $  &   $  0.101\pm0.09 $      &    90   &  $ 0.26\pm0.48 $  &   $-0.095\pm0.52 $    &    5      &    0.29  &                          Ra06,Al16 \\ 
            38628        Huya       & $  5.55\pm0.04 $  &   $ -0.152\pm0.03 $      &    45   &  $ 4.52\pm0.02 $  &   $ 0.078\pm0.01 $    &  104      &    0.10  &           BA03,Bo04,Ga16,SJ02,Al16 \\ 
            10370    Hylonome       & $  9.57\pm0.02 $  &   $  0.079\pm0.01 $      &     6   &  $ 8.98\pm0.04 $  &   $ 0.173\pm0.02 $    &    5      &    0.04  &                         RT99, BA03 \\ 
            28978       Ixion       & $  3.84\pm0.03 $  &   $  0.138\pm0.03 $      &    41   &  $ 3.25\pm0.04 $  &   $ 0.144\pm0.04 $    &    3      &    0.05  &                          Bo04,Al16 \\ 
                58534     Logos     & $  7.41\pm0.10 $  &   $  0.055\pm0.08 $      &     5   &  $ 6.72\pm0.10 $  &   $ 0.052\pm0.09 $    &    4      &    0.14   &                                 Al16 \\ 
               136472  Makemake     & $  0.00\pm0.01 $  &   $  0.206\pm0.01 $      &    53   &  $ 0.56\pm0.07 $  &   $-1.155\pm0.09 $    &    6      &    0.03  &                               Al16 \\ 
               52872    Okyrhoe     & $ 11.40\pm0.05 $  &   $ -0.013\pm0.01 $      &     7   &  $10.83\pm0.05 $  &   $ 0.020\pm0.01 $    &   12      &    0.07  &              BA03,Al16,\textbf{TW} \\ 
               90482      Orcus     & $  2.27\pm0.02 $  &   $  0.159\pm0.02 $      &    30   &  $ 1.87\pm0.03 $  &   $ 0.216\pm0.04 $    &    4      &    0.04  &                               Al16 \\ 
               49036     Pelion     & $ 10.89\pm0.08 $  &   $ -0.064\pm0.04 $      &     5   &  $10.35\pm0.07 $  &   $-0.029\pm0.04 $    &    6      &    0.14   &                          BA03,Al16 \\ 
               5145      Pholus     & $  7.46\pm0.31 $  &   $  0.152\pm0.15 $      &    10   &  $ 6.80\pm0.29 $  &   $ 0.111\pm0.13 $    &   16      &    0.60  &              BA03,Al16,\textbf{TW} \\ 
		\hline
	\end{tabular}                                        
\end{table*}
\addtocounter{table}{-1}
\begin{table*}
	\caption{Absolute Magnitudes}
	\begin{tabular}{|l|l|l|l|l|l|l|l|l|}  
\hline
                   Object           &        $H_V$      &     $\beta_V$            &  $N_V$  &     $H_R$         &        $\beta_R$      &  $N_R$    & $\Delta m$   &          Ref. \\	
                                    &        (mag)      & (mag per degree)         &         &      (mag)        &    (mag per degree)   &           &           &                   \\
\hline                                    
               50000     Quaoar     & $  2.77\pm0.25 $  &   $  0.116\pm0.22 $      &    45   &  $ 2.19\pm0.23 $  &   $ 0.047\pm0.24 $    &    8      &    0.30  &          Ba06,Al16 \\ 
               120347   Salacia     & $  3.51\pm0.06 $  &   $  0.665\pm0.04 $      &     9   &  $ 3.89\pm0.03 $  &   $-0.153\pm0.04 $    &   10      &    0.03  &   Al16, \textbf{TW} \\ 
               90377      Sedna     & $  1.56\pm0.01 $  &   $  0.640\pm0.04 $      &     9   &  $ 1.04\pm0.00 $  &   $ 0.166\pm0.00 $    &  157      &    0.02  &      Ra07,Pe10,Al16 \\ 
               79360 Sila-Nunam     & $  5.57\pm0.22 $  &   $  0.095\pm0.20 $      &     6   &  $ 4.91\pm0.16 $  &   $ 0.132\pm0.20 $    &    5      &    0.22  &           SJ02,Al16 \\ 
               32532    Thereus     & $  9.44\pm0.12 $  &   $  0.063\pm0.02 $      &    69   &  $ 9.00\pm0.30 $  &   $ 0.055\pm0.05 $    &   17      &    0.34  &      Je15,BA03,Al16 \\ 
               42355     Typhon     & $  7.66\pm0.02 $  &   $  0.128\pm0.01 $      &    22   &  $ 7.12\pm0.11 $  &   $ 0.138\pm0.04 $    &    5      &    0.07  &                Al16 \\ 
               174567     Varda     & $  3.97\pm0.04 $  &   $ -0.441\pm0.06 $      &    10   &  $ 3.24\pm0.04 $  &   $-0.024\pm0.06 $    &   10      &    0.06  &   Al16, \textbf{TW} \\ 
               20000     Varuna     & $  3.96\pm0.23 $  &   $  0.103\pm0.24 $      &    30   &  $ 3.29\pm0.21 $  &   $ 0.171\pm0.27 $    &   20      &    0.50  &     Hi05,Be06, Al16 \\ 	          
\hline
	\end{tabular}
	\caption[]{ 
    {\bf TW} = This work,
        Al16 = Alvarez-Candal et al. (2016),
        Ba00 = Barucci et al. (2000),
        Bo01 = Boehnhardt et al. (2001),
        Bo02 = Boehnhardt et al. (2002),     
        Bo04 = Boehnhardt et al. (2004),
        Ca06 = Carry et al. (2012),
        Ca12 = Carry et al. (2012),
        DM09 = Carraro et al. (2006),
        Do01 = Delsanti et al. (2001),
        Ga16 = Galiazzo et al. (2016),
        GH01 = Gil-Hutton et al. (2001),
        Hi05 = Hicks et al. (2005),
        Je15 = Jewitt et al. (2015),
        JL01 = Jewitt et al. (2001),
        Mu04 = Mueller et al. (2004),
        Px04 = Peixinho et al. (2004),
        Pe13 = Perna et al. (2013),
        Ra07 = Rabinowitz et al. (2007),  
        Ro05 = Rousselot et al. (2005),
        Se02 = Sekiguchi et al. (2002), 
        SJ02 = Sheppard et al. (2002), 
        Sn10 = Snodgrass et al. (2010),
        SS09 = Santos-Sanz et al. (2009), 
        Th12 = Thirouin et al. (2012),
        TR03 = Tegler \& Romanishin (2003). 
        } 
\end{table*}

\bsp	
\label{lastpage}

\begin{thebibliography}{99}
\bibitem[Alvarez-Candal et al.(2016)]{alvarez-candal16} Alvarez-Candal, A., Pinilla-Alonso, N., Ortiz, J.~L., et al.\ 2016, \aap, 586, A155 
\bibitem[Barucci et al.(1999)]{baruc00}Barucci, M.~A., Romon, J., Le Bras, A., Fulchignoni, M., \& Tholen, D.\ 1999, AAS/Division for Planetary Sciences Meeting Abstracts \#31, 31, 23.04 
\bibitem[Barucci et al.(2005)]{baruc05}Barucci, M.~A., Cruikshank, D.~P., Dotto, E., et al.\ 2005, Bulletin of the American Astronomical Society, 37, 56.01 
\bibitem[Barucci et al.(2011)]{baruc11}Barucci, M.~A., Alvarez-Candal, A., Merlin, F., et al.\ 2011, EPSC-DPS Joint Meeting 2011, 473 
\bibitem[Belskaya \& Shevchenko(2000)]{belsk00}Belskaya, I.~N., \& Shevchenko, V.~G.\ 2000, \icarus, 147, 94
\bibitem[Belskaya \& Shevchenko(2000)]{belsk10}Belskaya, I.~N., \& Shevchenko, V.~G.\ 2000, \icarus, 147, 94
\bibitem[Benecchi \& Sheppard(2013)]{Benecchi2013} Benecchi, S.~D., \& Sheppard, S.~S.\ 2013, \aj, 145, 124
\bibitem[Boehnhardt et al.(2001)]{boenh01} Boehnhardt, H., Tozzi, G.~P., Birkle, K., et al.\ 2001, \aap, 378, 653 
\bibitem[Boehnhardt et al.(2002)]{boenh02} Boehnhardt, H., Delsanti, A., Hainaut, O., et al.\ 2002, Asteroids, Comets, and Meteors: ACM 2002, 500, 47
\bibitem[Boehnhardt et al.(2004)]{boenh04} Boehnhardt, H., Bagnulo, S., Muinonen, K., et al.\ 2004, \aap, 415, L21
\bibitem[Dohnanyi(1971)]{Dohn71} Dohnanyi, J.~S.\ 1971, NASA Special Publication, 263.
\bibitem[Brown(2012)]{brown12} Brown, M.~E.\ 2012, Annual Review of Earth and Planetary Sciences, 40, 467
\bibitem[Carraro et al.(2006)]{carra06} Carraro, G., Maris, M., Bertin, D., \& Parisi, M.~G.\ 2006, \aap, 460, L39 
\bibitem[Carry et al.(2012)]{carra12} Carry, B., Snodgrass, C., Lacerda, P., Hainaut, O., \& Dumas, C.\ 2012, \aap, 544, A137 
\bibitem[Delsanti et al.(2001)]{delsa01} Delsanti, A.~C., Boehnhardt, H., Barrera, L., \& Hainaut, O.~R.\ 2001, Bulletin of the American Astronomical Society, 33, 12.04 
\bibitem[Doressoundiram et al.(2008)]{dores08} Doressoundiram, A., Boehnhardt, H., Tegler, S.~C., \& Trujillo, C.\ 2008, The Solar System Beyond Neptune, 91 
\bibitem[\protect\citeauthoryear{Farinella \& Davis}{1992}]{farinelladavis92} Farinella P., Davis D.~R., 1992, Icar, 97, 111
\bibitem[Galiazzo et al.(2016)]{gali16} Galiazzo, M., de la Fuente Marcos, C., de la Fuente Marcos, R., et al.\ 2016, \apss, 361, 212 
\bibitem[Gil-Hutton(2002)]{hutto02} Gil-Hutton, R.\ 2002, Planet. Space Science, 50, 57 
\bibitem[Gladman(2008)]{gladman08} Gladman, B.\ 2008, Bulletin of the American Astronomical Society, 40, 7.03 
\bibitem[Hicks et al.(2005)]{Hicks05} Hicks, M.~D., Simonelli, D.~P., \& Buratti, B.~J.\ 2005, \icarus, 176, 492
\bibitem[Hudson et al.(2008)]{hud08} Hudson, P.~K., Young, M.~A., Kleiber, P.~D., \& Grassian, V.~H.\ 2008, Atmospheric Environment, 42, 5991
\bibitem[Jewitt et al.(2001)]{Jewitt01J} Jewitt, D., Aussel, H., \& Evans, A.\ 2001, \nat, 411, 446 
\bibitem[Jewitt(2015)]{Jewitt15} Jewitt, D.\ 2015, \aj, 150, 201 
\bibitem[Ka{\v{n}}uchov{\'a}, et al.(2012)]{Kanuchova12} Ka{\v{n}}uchov{\'a}, Z., Brunetto, R., Melita, M., et al.\ 2012, \icarus, 221, 12.
\bibitem[Landolt(1992)]{lando92} Landolt, A.~U.\ 1992, \aj, 104, 340
\bibitem[Lacerda et al.(2014)]{lacer14} Lacerda, P., Fornasier, S., Lellouch, E., et al.\ 2014, \apjl, 793, L2
\bibitem[Luu \& Jewitt(1996)]{Luu1996} Luu, J. \& Jewitt, D.\ 1996, \aj, 112, 2310.
\bibitem[Muinonen et al.(2010)]{muin10} Muinonen, K., Belskaya, I.~N., Cellino, A., et al.\ 2010, \icarus, 209, 542
\bibitem[Muinonen, et al.(2010-b)]{muin-10b} Muinonen, K., Tyynel{\"a}, J., Zubko, E., et al.\ 2010, Earth, Planets, and Space, 62, 47.] 
\bibitem[Mueller et al.(2010)]{mulle10} Mueller, T.~G., Lellouch, E., Boehnhardt, H., et al.\ 2010, European Planetary Science Congress 2010, 668
\bibitem[Peixinho et al.(2012)]{peixe12} Peixinho, N., Delsanti, A., Guilbert-Lepoutre, A., Gafeira, R., \& Lacerda, P.\ 2012, \aap, 546, A86
\bibitem[Pike, et al.(2017)]{Pike2017} Pike, R.~E., Fraser, W.~C., Schwamb, M.~E., et al.\ 2017, AAS/Division for Planetary Sciences Meeting Abstracts 49, 504.12.
\bibitem[Rabinowitz et al.(2007)]{rabin07} Rabinowitz, D.~L., Schaefer, B.~E., \& Tourtellotte, S.~W.\ 2007, \aj, 133, 26
\bibitem[Roig, et al.(2008)]{Roig2008} Roig, F., Ribeiro, A.~O. \& Gil-Hutton, R.\ 2008, \aap, 483, 911
\bibitem[Rousselot et al.(2005)]{Rousselot05} Rousselot, P., Petit, J.-M., Poulet, F., \& Sergeev, A.\ 2005, \icarus, 176, 478
\bibitem[Santos-Sanz et al.(2009)]{Santos-Sanz09} Santos-Sanz, P., Ortiz, J.~L., Barrera, L., \& Boehnhardt, H.\ 2009, \aap, 494, 693 
\bibitem[Schaller \& Brown(2007)]{schal07} Schaller, E.~L., \& Brown, M.~E.\ 2007, European Planetary Science Congress 2007, 699
\bibitem[Sheppard \& Jewitt(2002)]{shepp02} Sheppard, S.~S., \& Jewitt, D.~C.\ 2002, \aj, 124, 1757
\bibitem[Shkuratov et al.(2002)]{shkur02} Shkuratov, Y., Ovcharenko, A., Zubko, E., et al.\ 2002, \icarus, 159, 396
\bibitem[Shkuratov(1994)]{shkur94} Shkuratov, Y.~G.\ 1994, Solar System Research, 28, 77.
\bibitem[Shkuratov(1994b)]{shkur94b} Shkuratov, Y.~G.\ 1994b, Astronomicheskii Vestnik, 28, 155.
\bibitem[Snodgrass et al.(2010)]{snodg10} Snodgrass, C., Carry, B., Dumas, C., \& Hainaut, O.\ 2010, European Planetary Science Congress 2010, 423
\bibitem[Strazzulla et al.(1991)]{strazz91} Strazzulla, G., Baratta, G.~A., Johnson, R.~E., \& Donn, B.\ 1991, \icarus, 91, 101
\bibitem[Tegler et al.(2003)]{Tegler03} Tegler, S.~C., Romanishin, W., \& Consolmagno, G.~J.\ 2003, \apjl, 599, L49 
\bibitem[Thirouin et al.(2010)]{thiro10} Thirouin, A., Ortiz, J.~L., Duffard, R., et al.\ 2010, VizieR Online Data Catalog, 352
\bibitem[Thirouin et al.(2012)]{thiro12} Thirouin, A., Ortiz, J.~L., Campo Bagatin, A., et al.\ 2012, \mnras, 424, 3156
\bibitem[Thompson et al.(1987)]{thompson87} Thompson, W.~R., Henry, T., Khare, B.~N., Flynn, L., \& Schwartz, J.\ 1987, \jgr, 92, 15083
\end{thebibliography}
\end{document}